\documentclass[a4paper,11pt]{article}
\pdfoutput=1 

\usepackage{jheppub} 
\usepackage{hyperref}
\usepackage[T1]{fontenc} 
\usepackage[normalem]{ulem}
\newcommand{\mpl}{M_{\text{pl}}}
\newcommand{\fnl}{f_{\text{NL}}}

\title{\begin{center}
		\LARGE{\textbf{Cosmological Collider Physics and the Curvaton}}
	\end{center}
}
\author{Soubhik Kumar}
\author{and Raman Sundrum}
\affiliation{Maryland Center for Fundamental Physics, Department of Physics,\\University of Maryland, College Park, MD 20742}
\emailAdd{soubhik@terpmail.umd.edu}
\emailAdd{raman@umd.edu}
\abstract{
Primordial non-Gaussianity signatures of extremely heavy particles are re-examined within a simple alternative to the standard inflationary paradigm, in which the primordial fluctuations and the inflationary spacetime expansion are sourced by two different fields. The curvaton scenario provides an example of this in which the distinct roles are played by the curvaton and the inflaton fields, respectively.  We study couplings of the curvaton to heavy particles with masses of order the inflationary Hubble scale, and show that they can lead to non-Gaussian signals orders of magnitude larger than those in standard inflation, consistent with explicit effective field theory control of inflationary dynamics. This brings various motivated particle physics signatures, such as loops of heavy gauge-charged scalars and fermions, within future observational reach. 
}

\begin{document}
	\hspace{30em} UMD-PP-019-04
	\maketitle
	\flushbottom
\section{Introduction}
An era of cosmic inflation (see \cite{Baumann:2009ds} for a review) in the very early universe is an attractive paradigm to explain the origin and the properties of the primordial density perturbations that eventually seed the temperature fluctuations in cosmic microwave background (CMB) and the inhomogeneities in large scale structure (LSS). The simplest models of cosmic inflation predict that the primordial fluctuations are adiabatic, scale invariant and approximately Gaussian---all three of these properties are verified to a good extent by the observations \cite{Akrami:2018odb}. Importantly, such properties can be tested with even better precision with upcoming surveys and this would allow us to get a more detailed picture of the inflationary era. In particular, any detection of primordial non-Gaussianity (NG) can give us crucial insight into physics at the inflationary Hubble scale $H$, which could lie well beyond the reach of terrestrial experiments. The time dependent inflationary spacetime can produce new particles with masses $\sim H$. If such particles can decay into inflatons, they can give rise to a distinctive non-analytic momentum dependence and angular dependence of the three (and higher)-point correlation function of the primordial fluctuations. Through a careful measurement of such dependencies, one can infer the mass and spin of such heavy particles---a rare opportunity to probe \textit{on-shell} physics at very high energy scales. This is the focus of the ``cosmological collider physics'' program \cite{Chen:2009zp,Baumann:2011nk,Assassi:2012zq,Chen:2012ge,Noumi:2012vr,Pi:2012gf,Gong:2013sma,Arkani-Hamed:2015bza,Lee:2016vti,Chen:2016nrs,Chen:2016uwp,Chen:2016hrz,Chen:2017ryl,Kehagias:2017cym,An:2017hlx,Kumar:2017ecc,Baumann:2017jvh}. For various recent ideas in this direction see, \cite{Chen:2018xck,An:2018tcq,Chen:2018cgg,Kumar:2018jxz,Arkani-Hamed:2018kmz,Li:2019ves,Wu:2019ohx,Alexander:2019vtb,Lu:2019tjj,Hook:2019zxa,Hook:2019vcn}.

To understand the prospects of detection of such heavy particle-induced NG, we can 
characterize the strength of NG by a dimensionless parameter, conventionally called $\fnl$ in the literature, whose definition will be given in sec. \ref{setup}. Current CMB constraints on $\fnl$ varies depending on the particular shape of NG under consideration and broadly, is given by $|\fnl|<\mathcal{O}(5-50)$ \cite{Akrami:2019izv}. Upcoming LSS experiments will improve this by having a precision $\sigma_{\fnl}\sim \mathcal{O}(1)$ \cite{Alvarez:2014vva} which will be useful to probe heavy particle-induced NG \cite{MoradinezhadDizgah:2017szk,MoradinezhadDizgah:2018ssw}.
The ultimate sensitivity in this regard will be provided by an only cosmic-variance-limited 21-cm experiment which can roughly achieve $\sigma_{\fnl}\sim 10^{-4}-10^{-3}$ \cite{Meerburg:2016zdz}. Thus we will consider $\fnl\sim 10^{-4}$ as the ultimate limiting strength of NG for observability.  Given this lower bound, with some rare exceptions \cite{Chen:2009zp}, many of the prime targets of the cosmological collider program such as, massive gauge bosons \cite{Chen:2016hrz,Chen:2016nrs,Chen:2016uwp,Kumar:2017ecc,Kumar:2018jxz}, charged scalars and fermions \cite{Chen:2016hrz,Chen:2016nrs,Chen:2016uwp,Kumar:2017ecc,Chen:2018xck,Hook:2019vcn,Hook:2019zxa}, Kaluza-Klein modes of the graviton \cite{Kumar:2018jxz} give rise to small, and sometimes even unobservable, strength of NG in the standard inflationary paradigm where the dynamics of inflation is \textit{explicitly} described in terms of scalar fields. The primary goal of the present work is to describe a simple, alternative paradigm, in which the above mentioned NGs are naturally orders of magnitude larger, and the associated targets can naturally be brought into the scope of the cosmological collider program. Let us first understand what suppresses such NG contributions in the standard paradigm.

\begin{figure}[h]
	\centering
	\includegraphics[width=0.25\linewidth]{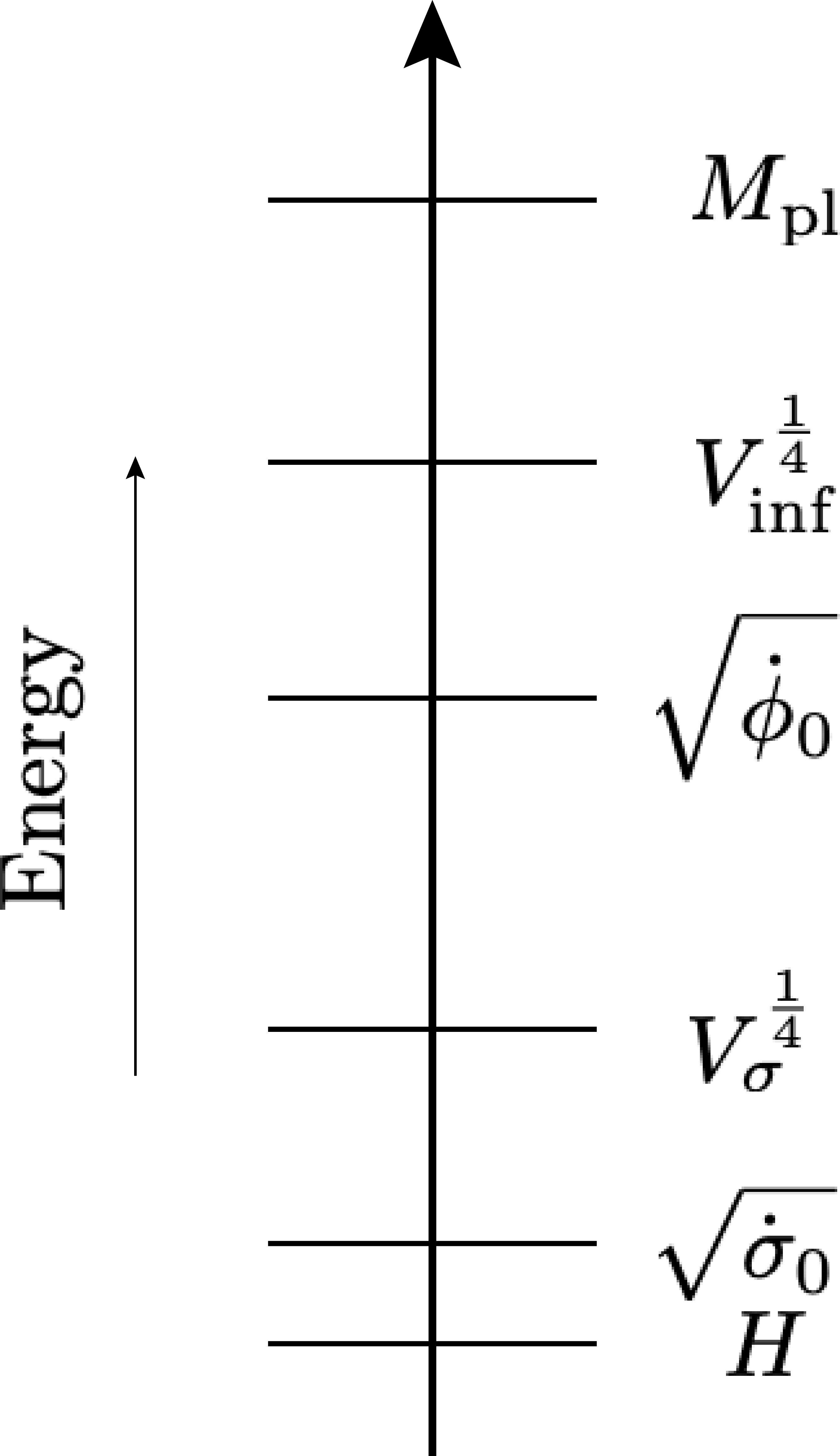}
	\caption{Various energy scales discussed in this work. $H$ and $\mpl$ are respectively, the inflationary Hubble scale and the Planck scale. $V_{\text{inf}}^{1/4}$ and $\sqrt{\dot{\phi}_0}$ are respectively the potential and kinetic energy scales of the inflaton field. Similarly, $V_{\sigma}^{1/4}$ and $\sqrt{\dot{\sigma}_0}$ are respectively the potential and kinetic energy scales of the curvaton field. A sample set of values of the above scales can be obtained using the benchmark parameter point given in eq. \eqref{benchmark}.}
	\label{fig:energy-scales}
\end{figure}

The inflaton ($\phi$) is a light field (with mass $\ll H$) during inflation. To ensure that it remains light in the presence of potentially large radiative corrections due to heavy states, one normally imposes a shift symmetry $\phi\rightarrow \phi+c$, with $c$ being a constant, which is broken only weakly by its potential. Under the restriction of such a shift symmetry, the dominant coupling of $\phi$ to a generic operator $\mathcal{O}$ is schematically described within an effective field theory (EFT) framework as,
\begin{align}\label{hd}
    \frac{1}{\Lambda_{\phi}^{2n+\text{dim}(\mathcal{O})-4}}(\partial\phi)^n\mathcal{O},
\end{align}
where $\text{dim}(\mathcal{O})$ is the scaling dimension of the operator $\mathcal{O}$, and $\Lambda_{\phi}$ is a scale by which the EFT description must break down. Strengths of such non-renormalizable couplings are thus characterized by inverse powers of $\Lambda_{\phi}$, and a smaller value of $\Lambda_{\phi}$ implies a larger coupling, leading to a larger strength of NG. So from an observational perspective, it is important to ask how small $\Lambda_{\phi}$ can be while our description still remains in theoretical control. Several possible choices of $\Lambda_{\phi}$ exist, with varying levels of conservatism and assumptions about the ultra-violet (UV) physics. A schematic representation of the relevant scales is shown in fig. \ref{fig:energy-scales}.

Due to quantum gravity effects any EFT description is expected to break down at $\mpl$ and thus $\Lambda_{\phi} \sim \mpl$ can be a reasonable choice, assuming no new physics comes in between $H$ and $\mpl$. With such a high value of $\Lambda_{\phi}$, the strength of NG is small, but can be observable in a few cases \cite{Chen:2009zp,Arkani-Hamed:2015bza,Kumar:2018jxz}. Stronger NG can be obtained by taking $\Lambda_{\phi}\gtrsim 
V_{\text{inf}}^{\frac{1}{4}}$ where $V_{\text{inf}}\sim H^2\mpl^2$ is the potential energy density that drives the inflationary expansion. Such a choice of $\Lambda_{\phi}$ is still conservative because $V_{\text{inf}}^{\frac{1}{4}}$ is the highest energy scale available during and after inflation. For example the reheat temperature, $T_R$, during the reheating stage at the end of inflation can be maximally $\sim V_{\text{inf}}^{\frac{1}{4}}$. Thus an EFT with $\Lambda_\phi\gtrsim V_{\text{inf}}^{\frac{1}{4}}$ is capable of describing the universe both during and after inflation.  Keeping this in mind, in this work we will take 
\begin{align}\label{lambdaPE}
\Lambda_\phi\gtrsim V_{\text{inf}}^{\frac{1}{4}}>250 H ,
\end{align}
where we have used the Planck constraint \cite{Akrami:2018odb} $H/\mpl < 2.7\times10^{-5}$ on tensor-to-scalar ratio. We will show later that the above restriction \eqref{lambdaPE} implies that the NG mediated by heavy particles in several scenarios of interest are quite small or even unobservable. 

We note that if one only demands theoretical control of the series of higher dimensional terms in the EFT which involve an expansion in $\frac{(\partial\phi)^2}{\Lambda_\phi^4}$, a weaker restriction of $\Lambda_{\phi}\gtrsim \sqrt{\dot{\phi}_0}\sim 60 H$
can be obtained \cite{Creminelli:2003iq}, where $\phi_0(t)$ denotes the homogeneous part of the inflaton field and we have used the fact that the scalar power spectrum implies $\frac{H^4}{\dot{\phi}_0^2}\approx 10^{-7}$ \cite{Akrami:2018odb}. However, such low $\Lambda_\phi$ does not explicitly capture the scalar field source of inflation, $V_{\text{inf}}$. We will not pursue this less conservative $\Lambda_\phi$ here.

Even more agnostically, the Goldstone effective theory of inflation \cite{Cheung:2007st}, which describes only the cosmological \textit{fluctuations} during inflation, but not the dynamics driving inflation, can avoid even the restriction $\Lambda_\phi\gtrsim \sqrt{\dot{\phi}_0}$. To ensure control over the description of just the fluctuations which probe energy scales $\sim H$, one need only have the EFT cut off $\Lambda_{\text{fluctuations}}\gtrsim H$. One can then have stronger couplings between the inflationary fluctuations and $H$-mass particles, leading to significantly larger NG than what is obtained by demanding \eqref{lambdaPE}.

This Goldstone description is completely agnostic about whether such low values of $\Lambda_{\text{fluctuations}}$ are consistent with higher scales such as $\sqrt{\dot{\phi}_0}$ and $V_{\text{inf}}^{1/4}$ since the latter scales, which control the homogeneous background, do not even appear in the Goldstone dynamics of inflationary fluctuations. The Goldstone description simply assumes that a suitable homogeneous inflationary background is given.
There do exist subtle mechanisms beyond the Goldstone description (e.g. \cite{Alishahiha:2004eh,Baumann:2011su}) which achieve compatibility between the two sets of scales. 
In this work we explore a very simple alternative where such a compatibility is readily obtained. This is achieved by having an explicit separation of the field degrees of freedom responsible for the inflationary background and for seeding the density fluctuations. 
We assume that along with the inflaton field, there exists a second light field $\sigma$, the ``curvaton'' \cite{Enqvist:2001zp,Lyth:2001nq,Moroi:2001ct}, whose role is to predominantly give rise to primordial fluctuations (today), whereas the role of the inflaton is reduced to just sourcing the background expansion (with subdominant fluctuations). Then it is completely consistent to have separate EFT cutoff scales for $\sigma$ and $\phi$, which automatically implies separate cutoffs for inflationary fluctuations and the homogeneous background. This can then naturally, but in a controlled way, lead to large NG if the EFT scale characterizing nonrenormalizable $\sigma$ interactions is parametrically smaller than the EFT cutoff of $\phi$ dynamics.
 
Concretely, one can have $\phi$ and $\sigma$ belong to two different sectors which are sequestered from each other (say, via having different locations in an extra dimension) with each having their own EFT cutoffs, $\Lambda_{\phi}$ and $\Lambda_{\sigma}$ respectively. Since $\sigma$ does not have to drive the background expansion, its energy density $V_\sigma$ can be $\ll V_{\text{inf}}$. Then an argument similar to the one leading to \eqref{lambdaPE}, will only imply
\begin{align}\label{lambdacurv}
\Lambda_{\sigma}\gtrsim V_{\sigma}^{\frac{1}{4}},
\end{align}
while still allowing $\Lambda_{\sigma}\ll V_{\text{inf}}^{\frac{1}{4}} <\Lambda_{\phi}$. Such a scenario with $\Lambda_{\sigma}\ll \Lambda_{\phi}$ can arise in several ways. One possibility is that the scale $\Lambda_{\sigma}$ could represent the masses of new mediator fields which couple to $\sigma$ and $H$-mass particles, but not to $\phi$. An example along this line will be studied in sec. \ref{curvatonparadigm}. Another possibility is that $\Lambda_{\sigma}$ could represent a compositeness/confinement scale for the $\sigma-$sector and/or heavy fields interacting with it. Via the AdS/CFT duality, this latter scenario is dual to an extra-dimensional set-up where $\phi$ and $\sigma$ are localized on two distinct ``branes'' and the extra-dimensional warping (redshifts) between the two branes explains why $\Lambda_{\sigma}\ll \Lambda_{\phi}$.

Since the two sectors are decoupled (up to gravitational effects) and can undergo separate reheating, $\Lambda_{\sigma}\ll V_{\text{inf}}^{\frac{1}{4}}$ does not lead to a break down of the EFT description for $\sigma$ at the end of inflation.
If the $H-$mass particles now couple to $\sigma$, such couplings need only be suppressed by $\Lambda_\sigma$, instead of $\Lambda_\phi$, and thus one can then have orders of magnitude bigger NG of primordial density fluctuations. Compared to the Goldstone EFT of inflation framework, this scenario can successfully describe inflationary and post-inflationary dynamics as will be discussed in sec. \ref{curvatonreview}, giving us an example of a controlled field theoretic scenario having potentially larger NG than the standard paradigm. 

This paper is organized as follows. In sec. \ref{setup}, we will set up the notation and discuss some essential aspects of the cosmological collider physics program. We will review the curvaton scenario and note the current set of observational constraints on it in sec. \ref{curvatonreview}. The detailed analysis of the EFT couplings of the heavy particles will then be carried out, both in the standard inflationary scenario, sec. \ref{inflatonparadigm}, and in the curvaton scenario, sec. \ref{curvatonparadigm}. Three types of NG contribution of heavy particles---tree-level effects of spontaneously broken charged scalars (Higgs bosons), and loop-level effects of unbroken charged scalars and charged Dirac fermions will be considered. We will discuss an issue of \textit{classical} tuning that arises in the inflationary paradigm and see that if we forego such tuning, the loop-mediated NG are unobservably small. While enhanced NG signals can be obtained with tuning, we will show that such tunings and enhancements are limited. The curvaton scenario will, on the other hand, have no such tunings but still gives rise to orders of magnitude larger NG compared to the standard inflationary scenario. 
In appendices \ref{scalarloop} and \ref{fermionloop} we will calculate the NG contributions explicitly to confirm those statements. We conclude in sec. \ref{conclude}. Throughout this paper we work in the $(-,+,+,+)$ sign convention for the spacetime metric.

As this paper was being completed, ref. \cite{Lu:2019tjj} appeared which discusses how Higgs fluctuations, different from inflaton fluctuations, can source primordial density perturbations through reheating via Higgs-modulated inflaton decay. The curvaton scenario we present here allows a larger enhancement of NG signals compared to ref. \cite{Lu:2019tjj} and also makes the robust prediction of the strength of a ``local'' type of NG, $\fnl^{\text{loc}}=-5/4$, absent in ref. \cite{Lu:2019tjj}. Nevertheless, there are some structural similarities between the present work and ref. \cite{Lu:2019tjj}.

\section{Observables and cosmological collider physics}\label{setup}
We denote the gauge invariant curvature perturbation, which will be defined below, by $\zeta$, and use the primed notation to denote its momentum space correlation functions, 
\begin{equation}
\langle \zeta(\vec{k}_1) \cdots \zeta(\vec{k}_n)\rangle  = (2\pi)^3\delta^3(\vec{k}_1+\cdots+\vec{k}_n)\langle \zeta(\vec{k}_1)\cdots\zeta(\vec{k}_n) \rangle^\prime.
\end{equation}
The power spectrum is denoted by,
\begin{align}
    P_k=\langle \zeta(\vec{k}) \zeta(-\vec{k})\rangle^\prime.
\end{align}
The dimensionless three point function parametrizing the strength of NG is defined as,
\begin{align}\label{F}
F\left(k_1,k_2,k_3\right)=\frac{\langle \zeta(\vec{k}_1) \zeta(\vec{k}_2)\zeta(\vec{k}_3)\rangle^\prime}{P_{k_1} P_{k_3}}.
\end{align}
The function $F$ defined above is in general momentum dependent and thus it is conventional in the literature to define an ``amplitude'' of NG in the equilateral limit $k_1=k_2=k_3$,
\begin{equation}
    \fnl=\frac{5}{18}F(k,k,k).
\end{equation}
Using the above definition of $\fnl$ one has a very rough estimate of the precision by which $\fnl$ can be measured in an only-cosmic-variance limited 21-cm experiment. Such a precision is controlled only by the number of modes $N_{\text{21-cm}}$ and is given by,
\begin{align}
    \sigma_{\fnl}\sim \frac{\langle\zeta^3\rangle}{\langle\zeta^2\rangle^{2}}\sim \frac{1}{\sqrt{N_{\text{21-cm}}}}\frac{1}{\zeta}.
\end{align}
Thus using the estimate $N_{\text{21-cm}}\sim 10^{16}$ \cite{Loeb:2003ya}, one has very roughly $\sigma_{\fnl}\sim 10^{-4}-10^{-3}$. 

We will be interested in the so-called ``squeezed limit'' of $F$ in eq. \eqref{F} for which $k_3\ll k_1\approx k_2$. In that case, $F$ becomes a function of only $\frac{k_3}{k_1}$. In particular, heavy fields with $M\sim H$, can mediate non-analytic momentum dependence of $F$ of the type,
\begin{equation}\label{smallf}
F_{\text{non-analytic}}\overset{k_3\ll k_1}{=}f_s(M)\left(\frac{k_3}{k_1}\right)^{\Delta_s(M)} P_s(\cos\theta)+ c.c.
\end{equation}
where $\theta=\hat{k}_1\cdot\hat{k}_3$. The functions $\Delta_s(M),f_s(M),P_s(\cos\theta)$ depend on the mass $M$ and spin $s$ of the heavy particle and can be calculated given its coupling to the inflaton. The prospect of extracting the mass and spin of the such heavy fields via measuring $\Delta_s(M)$ and $P_s$, forms the basis of cosmological collider physics. In the following, we will quantify the strength of NG by the absolute value $|f_s(M)|$.

While the time dependent inflationary spacetime readily produces particles with masses $\sim\mathcal{O}(H)$, production of heavier particles are ``Boltzmann suppressed'' with $f_s(M)\sim e^{-\pi M/H}$ for $M\gg H$. Furthermore, for $M\ll H$, $F$ becomes analytically dependent on $k_3$ so that the distinctive non-analytic, on-shell information characterizing heavy-particle mediation is no longer apparent. For example, for a scalar particle, $\Delta_0(M)=3/2\pm i\sqrt{M^2/H^2-9/4}\rightarrow 0 \text{ or }3$ for $M\ll H$. Thus the cosmological collider program operates most efficiently in a window of heavy masses around $\sim H$ to give us on-shell mass and spin information.

\section{Curvaton paradigm}\label{curvatonreview}
\subsection{Cosmological history}
We will now briefly review the cosmological history in the curvaton paradigm and emphasize some of the important differences between it and the standard inflationary paradigm. For more details on the curvaton paradigm, the reader is referred to the original papers \cite{Enqvist:2001zp,Lyth:2001nq,Moroi:2001ct}.

We will model the curvaton field, $\sigma$, as a pseudo Nambu-Goldstone boson (pNGB) whose shift symmetry is broken (softly) by a mass term $m\ll H$,
\begin{equation}\label{curv-pot}
    V_\sigma = \frac{1}{2}m^2\sigma^2.
\end{equation}
Any significant interaction term involving the curvaton and another field will need to respect a shift symmetry $\sigma\rightarrow \sigma+c$ with $c$ being some constant. Furthermore, for simplicity, we will assume that $\phi$ and $\sigma$ belong to two separate sectors sequestered from each other (say, by different locations in an extra-dimensional geometry) and ignore any interaction between them. Thus our model is specified by the lagrangian,

\begin{equation}
\mathcal{L} = -\frac{1}{2}(\partial\phi)^2 - V_{\text{inf}}(\phi) +\mathcal{L}^{\text{int}}_\phi(\partial_\mu\phi,\{\chi\})- \frac{1}{2}(\partial\sigma)^2 - \frac{1}{2}m^2\sigma^2 + \mathcal{L}^{\text{int}}_\sigma(\partial_\mu\sigma,\{\chi\}),
\end{equation}
where $V_{\text{inf}}(\phi)$ is the inflaton potential and $\mathcal{L}^{\text{int}}_{\phi(\sigma)}$ captures the shift-symmetric interactions of the inflaton (curvaton) with a collection of the other heavy fields $\{\chi\}$ that we will specify in sec. \ref{curvatonparadigm}. During inflation, the potential energy is dominated by $V_{\text{inf}}(\phi) \gg \frac{1}{2}m^2\sigma^2$ so that $\phi$ drives the inflationary expansion and $\sigma$ acts as a spectator field. 

To describe the fluctuations, we will split both the inflaton and the curvaton fields into homogeneous and fluctuating components: $\phi(t,\vec{x}) = \phi_0(t) + \delta\phi(t,\vec{x})$ and $\sigma(t,\vec{x}) = \sigma_0(t)+\delta\sigma(t,\vec{x})$. The equations of motion (EOM) for $\phi$ and $\sigma$ are decoupled (neglecting gravitational backreaction) and in particular, the homogeneous EOMs are given by \footnote{We will treat $\mathcal{L}^{\text{int}}_{\phi(\sigma)}$ in a perturbative manner so that they do not affect the free EOMs at the leading order.},
\begin{align}
    \ddot{\phi}_0+3H\dot{\phi}_0+V_{\text{inf}}'(\phi_0)=& 0,\\
    \ddot{\sigma}_0+3H\dot{\sigma}_0+m^2\sigma_0=& 0.
\end{align}
Assuming that the kinetic energy of the inflaton is much bigger than that of the curvaton, we get the standard relation, 
\begin{equation}
  \epsilon\equiv -\frac{\dot{H}}{H^2}\approx\frac{\dot{\phi}_0^2}{2H^2\mpl^2}. 
\end{equation}
Since $m^2\ll H^2$ and $\epsilon \ll 1$, the curvaton rolls very slowly along its potential, satisfying
\begin{equation}\label{sigmadot}
\dot{\sigma}_0 \approx -\frac{m^2}{3H}\sigma_0.
\end{equation}
Curvature fluctuations can be characterized by the gauge invariant quantity $\zeta$ defined by, 
\begin{equation}\label{zetadef}
    \zeta = -\psi -H \frac{\delta\rho}{\dot{\rho}_0}.
\end{equation}
In the above, $\psi$ is a spatial metric fluctuation appearing as,
\begin{equation}
    ds^2 = ((1-2\psi)\delta_{ij}+\cdots)a^2(t)dx^i dx^j+\cdots,
\end{equation}
and we have split the density $\rho(t,\vec{x})=\rho_0(t)+\delta\rho(t,\vec{x})$ into a homogeneous and a fluctuation part. For brevity, we have not explicitly written the other scalar, vector and tensor fluctuations.
Since the inflaton dominates the energy density during inflation, the curvature perturbation when the relevant momentum modes exit the horizon, $\zeta_{\text{exit}}$, is sourced only by $\delta\phi$ to a good approximation and thus, in a gauge in which $\psi=0$,
\begin{equation}
\zeta_{\text{exit}} \approx  -H \frac{\delta\phi}{\dot{\phi}_0}.
\end{equation}
One of the important features of the curvaton paradigm is that the fluctuations of the inflaton are subdominant to those of the curvaton, and in particular $\ll 10^{-5}$, the characteristic size of the observed primordial fluctuations. For example, for the benchmark set of parameters given in eq. \eqref{benchmark}, $\zeta_{\text{exit}}\sim \frac{H^2}{2\pi\dot{\phi}_0}\sim 10^{-6}$. However, significant curvature perturbations can get generated after the end of inflation since there is a second light field $\sigma$ during inflation and thus $\zeta$ need not necessarily be conserved on superhorizon scales \footnote{This is to be contrasted with single-field inflation where quite generally $\zeta$ remains conserved on superhorizon scales \cite{Wands:2000dp,Weinberg:2003sw}.}. Let us now see how this happens.

We assume that at the end of inflation, the inflaton reheats into a radiation bath largely decoupled from $\sigma$. In the meantime, $\sigma$ keeps rolling very slowly along its potential until the Hubble scale $\lesssim m$, following which $\sigma$ starts oscillating around its minimum and dilutes like matter. At such a point the content of the universe comprises of radiation coming from the inflaton decay, having energy density $\rho_{\text{rad}}$, and matter due to the curvaton energy density $\rho_\sigma$.
Thus using eq. \eqref{zetadef} and using the gauge $\psi=0$ the curvature perturbation after inflaton reheating can be written as,
\begin{equation}
\zeta = \frac{1}{3}\frac{\delta\rho_\sigma}{\rho_\sigma} f_\sigma + \frac{1}{4}\frac{\delta\rho_{\text{rad}}}{\rho_{\text{rad}}}(1-f_\sigma),
\end{equation}
where $f_\sigma = \frac{3\rho_\sigma}{3\rho_\sigma+4\rho_{\text{rad}}}$ is related to the energy density in the curvaton field compared to the radiation energy density $\rho_{\text{rad}}$. $\frac{\delta\rho_\sigma}{\rho_\sigma}$ and $\frac{\delta\rho_{\text{rad}}}{\rho_{\text{rad}}}$ are respectively fluctuations corresponding to the curvaton and the radiation (in the $\psi=0$ gauge), 
and they are conserved on super-horizon scales since the two fluids do not interact with each other, other than via gravity which is weak on these scales. Since the radiation bath originates from the inflaton decay, we will have $\frac{1}{4}\frac{\delta\rho_{\text{rad}}}{\rho_{\text{rad}}}=\zeta_{\text{exit}}$ which however is far subdominant in the curvaton scenario. Now, importantly since radiation dilutes faster than matter, assuming there is sufficient time between the start of curvaton oscillation and its decay, we will reach a stage at which $f_\sigma\approx 1$ when we can write,
\begin{equation}
\zeta \approx \frac{1}{3}\frac{\delta\rho_\sigma}{\rho_\sigma},
\end{equation}
which remains conserved on superhorizon scales subsequently. We assume all the relevant fluids during the later stage of evolution i.e. the SM photon, neutrinos, baryons and dark matter all originate from the decay of the curvaton. This way we do not generate any isocurvature fluctuations at a later stage. The differential evolution between $\sigma-$matter and radiation has converted the initial isocurvature fluctuations in the curvaton field into adiabatic ones.

We will now relate $\frac{\delta\rho_\sigma}{\rho_\sigma}$ to the quantum fluctuations of the curvaton field which will later help us to write expressions for NG of primordial density perturbations. 
Since the curvature perturbation is negligible at the end of inflation and we are assuming a mass-only potential for the curvaton, both $\sigma_0$ and $\delta\sigma(t,\vec{x})$ dilutes in an identical way so as to give \cite{Lyth:2001nq,Lyth:2002my},
\begin{equation}
\frac{\delta\rho_\sigma}{\rho_\sigma}=2\frac{\delta\sigma}{\sigma_0} =2 \frac{\delta\sigma}{\sigma_0}\vert_*
\end{equation}
where $*$ denotes the fact that fluctuations are evaluated at the epoch of horizon exit. This then finally gives,
\begin{equation}\label{zetafin}
\zeta_{\text{final}}\approx\zeta_\sigma = \frac{2}{3}\frac{\delta\sigma}{\sigma_0}\vert_*,
\end{equation}
which relates the final curvature perturbation in terms of the quantum fluctuations of the curvaton field. It is in this limit that the adiabatic curvaton fluctuations can be identified as the Goldstone mode for spontaneous time translation breaking in the Goldstone effective theory of inflation \cite{Cheung:2007st}, the inflaton fluctuations having become completely subdominant. 
Unless otherwise mentioned, in the following, we will omit the subscript in $\zeta_{\text{final}}$ and simply use $\zeta$ to denote the primordial density perturbations which act as ``initial'' conditions for the modes that subsequently re-enter the horizon after inflation. We are now in a position to note the present observational constraints on this paradigm.

\subsection{Observational constraints}
\paragraph{Scalar power spectrum.}

Due to the fact that $\sigma$ is a light spectator field during inflation,  its fluctuations $\delta\sigma$ acquire an approximately scale invariant spectrum. Thus the scalar power spectrum is given by, 
\begin{equation}\label{power}
\langle \zeta(\vec{k}) \zeta(-\vec{k})\rangle'= \frac{4}{9\sigma_0^2}\langle \delta\sigma(\vec{k}) \delta\sigma(-\vec{k})\rangle'
 = \frac{4}{9}\frac{H^2}{2\sigma_0^2k^3},
\end{equation}
where the r.h.s. is evaluated at the time of horizon exit $k=aH$ for a given $k-$mode.
The amplitude of scalar power spectrum from Planck data \cite{Akrami:2019izv} then implies, 
\begin{equation}\label{sigma0}
\frac{H}{\sigma_0} \approx 4.4\times 10^{-4}.
\end{equation}
\paragraph{Tilt of the scalar power spectrum.}
Defining $\Delta_\mathcal{\zeta} = \frac{k^3}{2\pi^2}\langle \zeta(\vec{k}) \zeta(-\vec{k})\rangle' = \frac{1}{9}\frac{H^2}{\pi^2\sigma_0^2},$
the tilt can be derived as,
\begin{equation}
\frac{d\ln\Delta_\mathcal{\zeta}}{d
\ln k} = -2\epsilon  +\frac{2}{3}\eta_\sigma.
\end{equation}
where $\eta_\sigma = \frac{m^2}{H^2}$ is fixed by the mass of the curvaton and $\epsilon\equiv-\frac{\dot{H}}{H^2}\approx\frac{\dot{\phi}_0^2}{2H^2 \mpl^2}$ is still determined by the homogeneous inflaton field. Planck data \cite{Akrami:2018odb} requires,
\begin{equation}
-2\epsilon  +\frac{2}{3}\eta_\sigma \approx -0.04.
\end{equation}
\paragraph{Tensor-to-Scalar ratio.}
The ratio of the power spectrum of tensor fluctuations to that of the scalar fluctuations, denoted by $r$, is given by,
\begin{equation}
r = \frac{\frac{8H^2}{\mpl^2}}{\frac{4H^2}{9\sigma_0^2}} =  \frac{18\sigma_0^2}{M_{\text{pl}}^2}.
\end{equation}
The upper bound $r<0.06$ from Planck data \cite{Akrami:2018odb} requires
\begin{align}
\sigma_0 < 0.06 \mpl.
\end{align}
\paragraph{Non-Gaussianity.}
A very stringent constraint on the curvaton paradigm comes from the upper bound on the ``local'' type of NG, defined as $\mathcal{\zeta} = \mathcal{\zeta}_g + \frac{3}{5}f_{\text{NL}}^{\text{loc}}\mathcal{\zeta}_g^2$, where $\zeta_g$ is a purely Gaussian field. Here the NG arises due to the fact that square of a Gaussian fluctuation is non-Gaussian. In the scenario when the curvaton dominates the energy density of the universe during before its decay, one can derive \cite{Lyth:2005fi,Bartolo:2003jx,Sasaki:2006kq}, 
\begin{equation}
    \mathcal{\zeta} = \frac{2}{3}\frac{\delta\sigma}{\sigma} - \frac{1}{3}\left(\frac{\delta\sigma}{\sigma}\right)^2.
\end{equation}
This has precisely the same form as the local type of NG defined above since $\delta\sigma$ is a Gaussian field, and in particular we have \footnote{Note that compared to \cite{Lyth:2005fi} our definition of $\fnl$ differs by an overall sign.},
\begin{equation}\label{fnlcurvaton}
f_{\text{NL}}^{\text{loc}} = -\frac{5}{4}.
\end{equation}
It should be noted that the above value of $f_{\text{NL}}^{\text{loc}}$ is parametrically larger than the slow-roll parameter suppressed $f_{\text{NL}}^{\text{loc}}$ in single-field inflationary models, as dictated by single-field consistency relations \cite{Maldacena:2002vr,Creminelli:2004yq} in the squeezed limit. Thereby in this curvaton scenario, even in the absence of heavy fields that will be considered below, a large $f_{\text{NL}}^{\text{loc}}$ is a tell-tale sign of beyond single-field inflationary dynamics. The above value of $f_{\text{NL}}^{\text{loc}}$ in eq. \eqref{fnlcurvaton} also serves as a crucial difference between the curvaton paradigm and Goldstone description of single-field inflation since $f_{\text{NL}}^{\text{loc}}$ is parametrically suppressed in the latter.


The above constraints can easily be satisfied. For example, one can choose a benchmark set of values:
\begin{equation}\label{benchmark}
    \sigma_0=5\times 10^{-3}\mpl;\hspace{1em} H = 2.2\times 10^{-6}\mpl; \hspace{1em}\epsilon=0.02;\hspace{1em}\eta_\sigma=10^{-3},
\end{equation}
to get $r=4.5\times 10^{-4}$. 
Although the robust prediction of  $f_{\text{NL}}^{\text{loc}}$ in eq. \eqref{fnlcurvaton} lies below the Planck upper bound on NG, quite excitingly, such a strength of NG will soon be tested by upcoming LSS observations \cite{Alvarez:2014vva}.

While it is true that the signal in the squeezed limit is dominated by the curvaton itself, the distinctive signaures of the heavy fields are imprinted in characteristic non-analytic ``oscillations'' in the squeezed limit as explained in eq. \eqref{smallf}. Multifield inflationary models having additional particles with masses $\ll H$ can not give such non-analytic momentum dependence. The observability of such oscillatory signals have been investigated in the literature in the context of single-field inflation, for example, in Refs. \cite{MoradinezhadDizgah:2017szk,MoradinezhadDizgah:2018ssw,Meerburg:2016zdz}. We expect that with some adaptations the above studies continue to be applicable in our case, but a detailed investigation lies beyond the scope of the present paper.

The fact that in the curvaton paradigm the background inflationary expansion and the (eventual) primordial density perturbations are sourced by two different fields, opens up an interesting possibility. In particular, during inflation both the kinetic and the potential energy stored in $\sigma$ can be much smaller than the kinetic and the potential energy stored in $\phi$, as can be checked by using the benchmark point in eq. \eqref{benchmark}. As will be explained below, this feature can allow significantly stronger coupling of some new degrees of freedom to $\sigma$ than to $\phi$ in light of  non-renormalizabality and shift symmetry of the couplings. We will illustrate this by considering the coupling of a charged scalar, with and without Higgsing, to both $\phi$ and $\sigma$, and the case of a Dirac fermion to both $\phi$ and $\sigma$.

\section{Charged heavy particles in the standard inflationary paradigm}\label{inflatonparadigm}
Since $\phi$ is a light field, we can model it as a pNGB in the low energy EFT, just like the curvaton, and impose a shift symmetry $\phi\rightarrow \phi+c$ which is broken only by its potential. This implies that the interaction of the inflaton will be characterized predominantly by a derivative expansion in $\frac{(\partial\phi)^2}{\Lambda_\phi^4}$ where $\Lambda_\phi$ is the EFT cutoff in the inflationary sector. As discussed in the introduction, theoretical control of such an expansion implies,
\begin{equation}\label{inflatonEFT}
\Lambda_\phi > \sqrt{\dot{\phi}_0} \sim  60H.
\end{equation}
A stronger restriction on $\Lambda_\phi$ can be placed if we demand that the EFT explicitly describes the scalar/gravity dynamics of inflation and reheating. All known descriptions of this refer to an inflaton potential as the source of inflationary expansion. Thus the control of such an EFT requires \eqref{lambdaPE},
\begin{align}\label{Vinf}
    \Lambda_\phi > V_{\text{inf}}^{\frac{1}{4}}>250 H.
\end{align}
In the following we will keep only the restriction in eq. \eqref{lambdaPE} in mind while considering the strengths of NG.

The leading coupling of the inflaton to a scalar field $\chi$, charged under some gauge/global symmetry group, is given by a dimension-6 operator 
\begin{align}\label{inf-higgs}
\mathcal{L}\supset\frac{1}{\Lambda_\phi^2}(\partial\phi)^2 \chi^\dagger \chi.
\end{align}
This term will also contribute to the mass of $\chi$ since,
$\frac{1}{\Lambda_\phi^2}(\partial\phi)^2 \chi^\dagger \chi\supset -\alpha\chi^\dagger \chi$ where $\alpha = \frac{\dot{\phi}_0^2}{\Lambda_\phi^2}$ is approximately constant in slow-roll inflation. In the presence of a ``bare'' mass $m_\chi$ and a quartic coupling $\lambda_\chi$, the lagrangian for $\chi$ then reads as,
\begin{align}
\mathcal{L}\supset -|\partial\chi|^2-(m_\chi^2+\alpha)\chi^\dagger \chi -\lambda_\chi (\chi^\dagger \chi)^2,
\end{align}
where the effective mass for $\chi$ is given by,
\begin{equation}\label{masstuning1}
m_{\chi,\text{eff}}^2=m_{\chi}^2+\alpha.
\end{equation}
Two scenarios arise depending on the sign of $m_{\chi,\text{eff}}^2$. 

\subsection{Higgs exchange in the broken phase}
We first discuss the case when $m_{\chi,\text{eff}}^2 < 0$, leading to a Higgsing of the symmetry. Due to spontaneous symmetry breaking, one can now have inflationary couplings which are linear in the heavy field. Consequently, one can have tree level processes that can mediate NG and since such processes do not have the usual $\sim\frac{1}{16\pi^2}$ loop suppression, the associated NG are more readily observable. We consider a $U(1)$ symmetry group for simplicity. Given the coupling in eq. \eqref{inf-higgs}, one can expand $\chi = (0, \frac{1}{\sqrt{2}}(v+\tilde{\chi}))$ in the unitary gauge, to read off the vertices necessary for tree level NG. In the above, $v$ and $\tilde{\chi}$ are respectively the VEV and fluctuations of the Higgs field. 

\begin{figure}[h]
	\centering
	\includegraphics[width=0.8\linewidth]{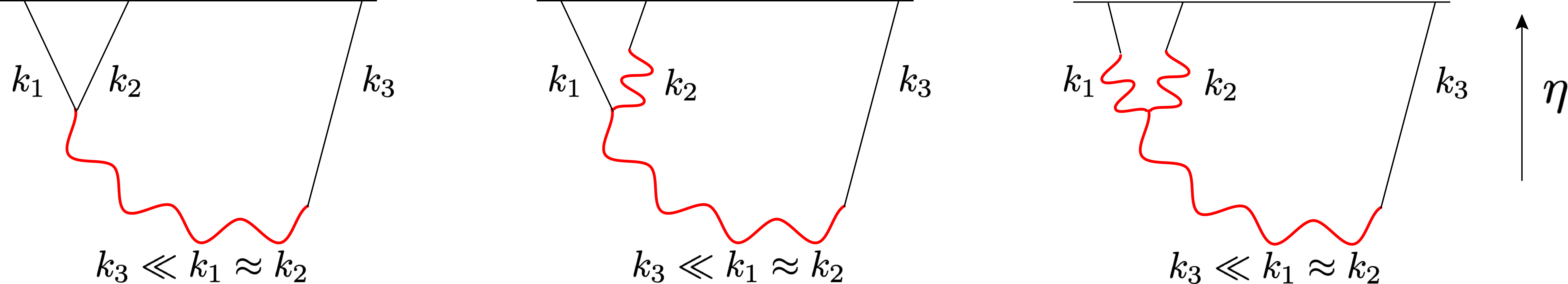}
	\caption{Massive Higgs mediated (in red) tree level ``in-in'' contributions to the inflaton (in black) three point function. Depending on the number of massive scalar propagators, these diagrams are labelled from left to right: (a) single exchange diagram, (b) double exchange diagram, (c) triple exchange diagram. $\eta$ denotes conformal time which ends at the end of inflation.}
	\label{fig:diagrams}
\end{figure}

The details have been discussed in \cite{Kumar:2017ecc} and the summary is that one can have three types of diagrams giving rise to NG as shown in fig. \ref{fig:diagrams}.

The rough strengths of NG, in terms of $f$ defined eq. \eqref{smallf}, corresponding to each of the diagrams are \cite{Kumar:2017ecc},
 \begin{align}\label{treehiggsng}
|f_{\chi,\text{ single, tree}}|\sim\frac{\rho_1^2}{H^2}; \hspace{2em} |f_{\chi,\text{ double, tree}}|\sim\frac{\rho_1^2\alpha}{H^4}; \hspace{2em} |f_{\chi,\text{ triple, tree}}|\sim\frac{\rho_1^2\alpha}{H^4},
 \end{align}
where $\rho_1 = \frac{2\alpha v}{\dot{\phi}_0}$ denotes the quadratic mixing between $\phi$ and $\tilde{\chi}$. 

For observable strengths of NG, we need the masses of the heavy particles $\sim H$, otherwise the cosmological production of the massive particle will be severely Boltzmann suppressed. Eq. \eqref{masstuning1} then implies that in the absence of any \textit{classical} tuning between $m_\chi^2$ and $\alpha$ we need to have both $\alpha\sim m_\chi^2\sim H^2$, in which case all three diagrams will give similar strength of NG, $|f_{\chi,\text{ natural, tree}}|\sim \rho_1^2/H^2$. Since a quadratic mixing between $\phi$ and $\tilde{\chi}$ also gives rise to a correction to the scalar power spectrum $k^3 P_k\sim \frac{H^4}{\dot{\phi}_0^2}(1+\mathcal{O}(\rho_1^2/H^2))$, we will require $\rho_1^2/H^2\lesssim 0.1$ for perturbativity of such corrections. Then we see that NG contributions are given by,
\begin{align}
    |f_{\chi,\text{ natural, tree}}|\lesssim 0.1.
\end{align}
The natural choice of $\alpha\sim H^2$ implies $\Lambda > V_{\text{inf}}^{\frac{1}{4}} > \sqrt{\dot{\phi}_0}$, also ensuring a controlled EFT description.

From eq. \eqref{treehiggsng} it is clear that by choosing a larger value of $\alpha$ and consequently fine tuning it against $m_\chi^2$ to obtain $m_{\chi,\text{eff}}^2\sim H^2$, a larger strength of NG can be obtained. However, one can not do this tuning to more than a percent level since the (slow) time evolution of $\alpha = \frac{\dot{\phi}_0^2}{\Lambda^2}$ will generically push $m_{\chi,\text{eff}}^2$ away from its tuned value $\sim H^2$ in a few Hubble times.

\subsection{Charged scalar exchange in the symmetric phase}
Here we assume $m_{\chi,\text{eff}}^2>0$ so that there is no spontaneous symmetry breaking and $\chi$ mediated NG appear only via loop diagrams.
The coupling to the inflaton is described by the same operator as above, namely $\frac{1}{\Lambda_\phi^2}(\partial\phi)^2 \chi^\dagger \chi$. With the symmetry being unbroken the relevant couplings between the inflaton and $\chi$ are given by,
\begin{equation}\label{inf-Higgs-coupling}
  \mathcal{L}_{\phi-\chi}\supset\frac{1}{\Lambda_\phi^2}(\partial\phi)^2 \chi^\dagger \chi = \left(-\alpha - \frac{2\alpha}{\dot{\phi}_0} \dot{\delta\phi}+\frac{\alpha}{\dot{\phi}_0^2}(\partial(\delta\phi))^2\right)\chi^\dagger \chi,
\end{equation}
with an effective $\chi$ mass given by eq. \eqref{masstuning1} and $\alpha=\frac{\dot{\phi}_0^2}{\Lambda_{\phi}^2}$.

\begin{figure}[h]
	\centering
	\includegraphics[width=0.7\linewidth]{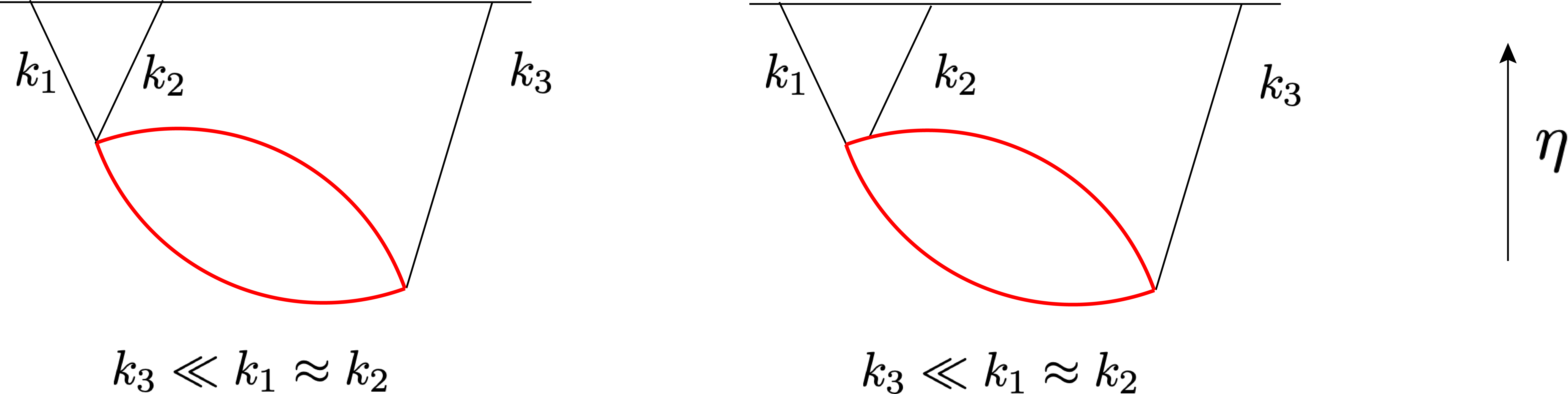}
	\caption{Massive charged particle mediated (in red) loop level ``in-in'' contributions to the inflaton (in black) three point function. Depending on the number of massive charged particle propagators, these diagrams are labelled from left to right: (a) double exchange diagram, (b) triple exchange diagram. $\eta$ denotes conformal time which ends at the end of inflation.}
	\label{fig:scalar-loop}
\end{figure}

Based on the couplings given in eq. \eqref{inf-Higgs-coupling}, there are two loop diagrams that can contribute to a three point function which we list in fig. \ref{fig:scalar-loop}.
The associated NG can be estimated, in terms of $f$ defined eq. \eqref{smallf}, as
 \begin{align}\label{loopscalar}
|f_{\chi,\text{ double, loop}}|\sim\frac{1}{16\pi^2}\frac{\alpha^2}{\dot{\phi}_0^2}; \hspace{2em} 
|f_{\chi,\text{ triple, loop}}|\sim\frac{1}{16\pi^2}\frac{\alpha^3}{H^2\dot{\phi}_0^2}.
 \end{align}
The above discussion of the ``classical'' tuning also applies here and in the natural case i.e. when $\alpha\sim H^2$, eq. \eqref{loopscalar} gives,
\begin{equation}\label{brokenscalrinf}
	|f_{\chi,\text{ natural, loop}}|\sim \frac{1}{16\pi^2}\frac{H^4}{\dot{\phi}_0^2}\sim 10^{-9},
\end{equation}
where we have used the fact that $\frac{H^4}{\dot{\phi}_0^2}\sim 10^{-7}$. Such a strength of NG is unobservably small. 
Note already with $\alpha\sim H^2$, $m_{\chi,\text{eff}}^2$ receives $\mathcal{O}(1)$ ``contamination'' from the inflationary background and a measurement of $m_{\chi,\text{eff}}^2$ via NG would not have given us the underlying value of the ``pure'' mass $m_\chi^2$. For $\alpha \ll H^2$, such a contamination is small but the strength of NG becomes even smaller. 

\subsection{Charged Dirac fermion}
We will consider a charged Dirac fermion coupled to the inflaton via a dimension-7 operator
$\frac{1}{\Lambda_\phi^3}(\partial\phi)^2 \bar{\Psi}\Psi$. A dimension-5 operator of type $\frac{1}{\Lambda_\phi}\partial_\mu\phi\bar{\Psi}\gamma^\mu\Psi$ can be eliminated by integration by parts and current conservation if the $\Psi$ couplings respect a $U(1)$ symmetry. 

There could exist another dimension-5 operator involving the axial current, $\frac{\partial_\mu\phi\bar{\Psi}\gamma^\mu\gamma^5\Psi}{\Lambda_\phi}$ if it is not forbidden by parity. Such a coupling is special since it gives rise to an effective ``chemical potential'' $\lambda = \dot{\phi}_0/\Lambda_\phi$ for the fermion $\Psi$ once the inflaton is set to its background value \cite{Chen:2018xck,Hook:2019zxa,Hook:2019vcn}. It can help production of $\Psi$ even when the fermion mass, $m_\Psi$ is somewhat heavier than $H$ without paying significant Boltzmann suppression. 
However, if we impose the restriction in \eqref{lambdaPE} i.e. $\lambda\lesssim 15 H$, and demand theoretical control of the calculation of NG in the squeezed limit, which forces $\frac{k_3}{k_1}<\frac{H}{\lambda}$ \cite{Hook:2019zxa}, we find $F\lesssim \text{few}\times 10^{-4}$ for the function defined in eq. \eqref{F}. We will see in the next section that the curvaton scenario will allow a larger strength of NG with just the analog of the dimension-7 operator defined above. Furthermore, in the regime where there is a substantial NG signal due to $\lambda$, the inflationary background also significantly contaminates the ``pure'' fermion mass $m_\Psi$, so that the non-analytic signatures are predominantly sensitive to $\lambda$, not $m_\Psi$.  Because of these reasons we will not consider the dimension-5 operator $\frac{\partial_\mu\phi\bar{\Psi}\gamma^\mu\gamma^5\Psi}{\Lambda_\phi}$ further.


In the coupling $\frac{1}{\Lambda_\phi^3}(\partial\phi)^2 \bar{\Psi}\Psi$, the VEV of the inflaton will give a contribution to the fermion mass as for charged scalars, and we can write the relevant couplings as,
\begin{equation}\label{inf-fermion-coupling}
    \mathcal{L}\supset\frac{1}{\Lambda_\phi^3}(\partial\phi)^2 \bar{\Psi}\Psi = \left(-\beta - \frac{2\beta}{\dot{\phi}_0} \dot{\delta\phi}+\frac{\beta}{\dot{\phi}_0^2}(\partial(\delta\phi))^2\right)\bar{\Psi}\Psi,
  \end{equation}
where $\beta = \frac{\dot{\phi}_0^2}{\Lambda_\phi^3}$. Thus the effective fermion mass becomes,
\begin{equation}\label{masstuning2}
m_{\Psi,\text{eff}}=m_{\Psi}+\beta.
\end{equation}
Naturalness requires $\beta\sim H$. Furthermore, even if we choose to tune, the requirement of $\Lambda_\phi > \sqrt{\dot{\phi}_0}$ already implies,
\begin{equation}
    \beta < \sqrt{\dot{\phi}_0}.
\end{equation}
Focusing on the natural case we can estimate the strength of NG for a fermion-mediated loop process given in fig. \ref{fig:scalar-loop}, where the internal lines now represent $\Psi$, in terms of $f$ defined eq. \eqref{smallf},
\begin{align}\label{loopfermion}
|f_{\Psi,\text{ double, loop}}|\sim\frac{1}{16\pi^2}\frac{\beta^2H^2}{\dot{\phi}_0^2}; \hspace{2em} 
|f_{\Psi,\text{ triple, loop}}|\sim\frac{1}{16\pi^2}\frac{\beta^3H}{\dot{\phi}_0^2}.
 \end{align}
We see in the natural case i.e. $\beta \sim H$ such NG are again unobservably small with
\begin{equation}\label{fermioninf}
	|f_{\Psi,\text{ natural, loop}}|\sim 10^{-9}.
\end{equation}
As in the case of the charged scalar, if we demand the ``contamination'' to $m_{\Psi,\text{eff}}$ from the inflationary background to be small, the NG becomes even smaller.

Having discussed the case with inflationary couplings, let us see how things change when the heavy charged scalar and fermion fields are coupled to a curvaton instead of the inflaton as the dominant source of fluctuations. 
\section{Charged heavy particles in the curvaton paradigm}\label{curvatonparadigm}
In the inflationary scenario, it is the same field $\phi$ that drives the inflationary expansion and also sources the observed primordial fluctuations. That is why the very high scales such as $\dot{\phi}_0^{1/2}$ and $V_{\text{inf}}^{1/4}$ inversely bound the strength of EFT couplings, making the associated NG small. However, in the curvaton scenario, $\sigma$ sources the observed fluctuations and $\phi$ drives the inflationary expansion. In particular, assuming $\phi$ and $\sigma$ belong to two different sectors, sequestered from each other, the scales $\dot{\phi}_0^{1/2}$ and $V_{\text{inf}}^{1/4}$ need not even be relevant for the couplings between $\sigma$ and heavy fields. The only relation that is relevant for control of the curvaton EFT is,
eq. \eqref{lambdacurv},
\begin{align}
    \Lambda_\sigma > V_\sigma^{\frac{1}{4}},
\end{align}
where $V_\sigma$ is the energy density in the curvaton. 
To see how big $V_\sigma$ can be we write,
\begin{equation}\label{Vsigma}
V_\sigma= \frac{1}{2}m^2\sigma_0^2 \approx  2.5\times 10^6\times \eta_\sigma H^4,
\end{equation}
where we have used eq. \eqref{sigma0} and $\eta_\sigma = \frac{m^2}{H^2}$.
We then see that for the benchmark choice of $\eta_\sigma\sim 10^{-3}$ discussed in sec. \ref{curvatonreview} allows us to have using eq. \eqref{lambdaPE},
\begin{align}
\Lambda_\phi\gtrsim 250 H \gg \Lambda_\sigma \gtrsim V_\sigma^{1/4}\sim 10 H
\end{align}
This relation $\Lambda_\phi \gg \Lambda_\sigma \gtrsim V_\sigma^{1/4} $ plays a central role in giving significantly larger NG in the curvaton paradigm while also ensuring theoretical control of the set-up.
Compared to the standard inflationary paradigm, in the curvaton scenario there are no \textit{classical} tunings so long as we make sure $\dot{\sigma}_0\lesssim H^2 < \Lambda_\sigma^2$. As an example, the contribution of the curvaton to the mass term of the charged scalar field $\chi$, 
$\frac{(\partial\sigma)^2}{\Lambda_\sigma^2}\chi^\dagger\chi\supset -\frac{\dot{\sigma}_0^2}{\Lambda_\sigma^2}\chi^\dagger\chi$ remains small for the above choices of $\dot{\sigma}_0$ and $\Lambda_\sigma$ and correspondingly there is no ``contamination'' to the ``pure'' scalar mass $m_\chi$.

\paragraph{Constraints from curvaton self-interaction mediated NG.}
Given the derivative expansion, we can also write a term $\frac{1}{\Lambda_\sigma^4}(\partial\sigma)^4$. This will contribute to both bispectrum and trispectrum. For bispectrum, the relevant couplings are, 
\begin{equation}
\frac{1}{\Lambda_\sigma^4}(\partial\sigma)^4 = 4\frac{\dot{\sigma}_0}{\Lambda_\sigma^4}\dot{\delta\sigma}^3 -4\frac{\dot{\sigma}_0}{\Lambda_\sigma^4}\frac{1}{a(t)^2}\dot{\delta\sigma}(\partial_i\delta\sigma)^2+\cdots.
\end{equation}
From such a coupling we can do a naive estimate of curvaton self-interaction mediated NG, in terms of $f$ defined eq. \eqref{smallf},
\begin{equation}
    |f_{\text{curvaton, self-int.}}|\sim \frac{\dot{\sigma}_0 \sigma_0 H}{\Lambda_\sigma^4}.
\end{equation}
Thus with $\dot{\sigma}_0\sim H^2, \Lambda_\sigma\sim 4 H$, a choice which will be justified below, and using eq. \eqref{sigma0} we get, $f_{\text{curvaton, self- int.}}\sim 10$. This is below the current upper bound on orthogonal and equilateral type of NG which is the kind of NG induced by the above self-interaction. Doing a more careful calculation shows that $f_{\text{curvaton self int.}}$ is actually smaller than the above crude estimate, so that a choice of $\Lambda_\sigma\gtrsim 4 H$ is more than sufficient to avoid current constraints. With such a choice of $\Lambda_\sigma$, the trispectrum is also smaller than the current bound \cite{Akrami:2019izv}.
\paragraph{Cut-off of the effective theory and field range of the curvaton.}
With the above choice of $\eta_\sigma\sim 10^{-3}$, the requirement $\Lambda_\sigma > V_\sigma^{\frac{1}{4}}$ reduces to $\Lambda_\sigma \gtrsim  10 H$. However, even with $\Lambda_\sigma\sim 10 H$, we do still need $\sigma_0\gg \Lambda_\sigma$ to satisfy eq. \eqref{sigma0}. Thus one can ask whether it is problematic to have curvaton field range much bigger than the EFT cut-off, although both the derivative expansion
remains under full control and $\Lambda_{\sigma} > V_\sigma^{\frac{1}{4}}$. This is analogous to the problem of super-Planckian inflaton field range in high scale inflation models, given by the Lyth bound. Thus, one can borrow the mechanisms that are used to create large effective  field ranges, such as axion monodromy or multi-axion alignment. These mechanisms illustrate that having field ranges larger than the EFT cutoff can naturally emerge from controlled UV completions. In particular, following the bi-Axion mechanism \cite{Kim:2004rp,Choi:2014rja,Tye:2014tja,Ben-Dayan:2014zsa,Bai:2014coa,delaFuente:2014aca}, one can imagine having two axionic curvatons with potential,
\begin{equation}
V_{\text{curv}} = V_1\left(1-\cos\left(\frac{N \sigma_1}{f_1}+\frac{\sigma_2}{f_2}\right)\right) + V_2\left(1-\cos\left(\frac{\sigma_1}{f_1}\right)\right).
\end{equation}
With $V_1\sim V_2$, $f_1\sim f_2$ and $N\gg 1$, there is a heavier curvaton which gets stabilized at $\frac{N \sigma_1}{f_1}+\frac{\sigma_2}{f_2}\approx 0$ so that the effective light curvaton potential is given by,
\begin{equation}\label{curvpotfull}
    V_{\text{eff}} = V_2\left(1-\cos\left(\frac{\sigma_2}{Nf_2}\right)\right).
\end{equation}
We see that the light curvaton $\sigma_2$ has an effective field space $\sim Nf_2$ that is parametrically bigger than the fundamental field space $f_2$ and we can consistently have $Nf_2\gg\Lambda_\sigma\gtrsim f_2$ for a sufficiently large $N$. This UV completion then serves as a proof-of-principle that it is consistent to have field ranges that are significantly bigger than the EFT cutoff scales. The light curvaton $\sigma_2$ can still interact with heavy fields of interest with suppressions given by $\Lambda_\sigma\sim f_2$ rather than $Nf_2$, so that we can have stronger couplings leading to significant NG. Expanding eq. \eqref{curvpotfull} around its minima gives rise to the approximately mass-only potential for the curvaton considered in eq. \eqref{curv-pot} with $\sigma_2=\sigma$ and $m^2 = \frac{V_2}{N^2f_2^2}$.

\paragraph{Mediators and stronger EFT couplings.} 
The curvaton will couple to heavy particles via higher dimensional operators and hence the NG will be proportional to multiple powers of $\frac{1}{\Lambda_\sigma}$, instead of multiple powers of $\frac{1}{\Lambda_\phi}$ as in the inflationary scenario. As we have discussed above, the strength of NG in the curvaton scenario can therefore be much stronger since $\Lambda_\sigma\ll\Lambda_\phi$ is consistent with EFT control. However, the restriction of $\Lambda_\sigma> 10 H$ still corresponds to somewhat suppressed NG. We will now show that in the presence of heavier ``mediator'' particles, the effective scale of $\Lambda_\sigma$ can be brought down from $\sim 10 H$ and we can obtain even stronger NG.

For example, one can have the following coupling between the curvaton, the mediator $\Sigma$ and the scalar $\chi$:
\begin{equation}
   \frac{1}{\Lambda_\sigma}(\partial\sigma)^2\Sigma+\mu_\Sigma\Sigma \chi^\dagger \chi. 
\end{equation}
Upon integrating out the mediator $\Sigma$, we can get an effective dimension-6 operator,
\begin{equation}
  \frac{\mu_\Sigma}{M_\Sigma^2\Lambda_\sigma}(\partial\sigma)^2 \chi^\dagger \chi,
\end{equation}
which implies an effective cutoff,
\begin{equation}
    \Lambda_{\sigma,\text{eff}}^2 = \frac{M_\Sigma^2 \Lambda_\sigma}{\mu_\Sigma}.
\end{equation}
As an example, for $M_\Sigma = 3H, \mu_\Sigma=6H$ and $\Lambda_\sigma = 10 H$, one gets $\Lambda_{\sigma,\text{eff}} \approx 4 H < \Lambda_\sigma$.
A similar procedure can be repeated for fermionic couplings by starting with,
\begin{equation}
 \frac{1}{\Lambda_\sigma}(\partial\sigma)^2\Sigma+y\Sigma \bar{\Psi}\Psi
\end{equation}
to get (for $y=1$),
\begin{equation}
    \Lambda_{\sigma,\text{eff}}^3 = M_\Sigma^2 \Lambda_\sigma.
\end{equation}
This effective cutoff is again smaller than $\Lambda_\sigma$ for the same choice of $M_\Sigma$ and $\Lambda_\sigma$. 

To summarize, demanding theoretical control of the curvaton derivative expansion and $\Lambda_\sigma > V_\sigma^{\frac{1}{4}}$ in the presence of somewhat heavy mediators, gives us $\Lambda_{\sigma,\text{eff}}\gtrsim 4H$. In the following, we will give parametric estimates of NG in terms of $\Lambda_{\sigma,\text{eff}}$ but for numerical results we will take
$\Lambda_{\sigma,\text{eff}} = 4H$. For brevity, in the rest of the paper we will denote $\Lambda_{\sigma,\text{eff}}$ by just $\Lambda_{\sigma}$.

\subsection{Higgs exchange in the broken phase}
The coupling of the curvaton with the charged scalar is given by the dimension-6 operator,
\begin{equation}\label{curv-chi-tree}
    \mathcal{L}_{\phi-\chi}\supset\frac{1}{\Lambda_\sigma^2}(\partial\sigma)^2 \chi^\dagger \chi .
\end{equation}
Expanding around the correct vacuum, $\chi=(0,\frac{1}{\sqrt{2}}(\tilde{\chi}+v))$ as in sec. \ref{inflatonparadigm}, we can get the relevant terms coupling $\tilde{\chi}$ to $\sigma$,
\begin{equation}\label{treescalar2}
\mathcal{L}_{\sigma-\chi}\supset\frac{1}{\Lambda_\sigma^2}(\partial\sigma)^2 \chi^\dagger\chi-\lambda_\chi(\chi^\dagger \chi)^2 \supset \left(-\frac{2\dot{\sigma}_0v}{\Lambda_\sigma^2}\dot{\delta\sigma}\tilde{\chi}-\frac{\dot{\sigma}_0}{\Lambda_\sigma^2}\dot{\delta\sigma}\tilde{\chi}^2+\frac{v}{\Lambda_\sigma^2}(\partial\delta\sigma)^2\tilde{\chi}-\lambda_\chi v \tilde{\chi}^3\right)+\cdots.
\end{equation}
The above couplings give NG mediated by diagrams given in fig. \ref{fig:diagrams} where the external legs represent fluctuations of the curvaton instead of the inflaton. The leading contribution comes from the single exchange diagram for which one can roughly estimate the NG, in terms of $f$ defined eq. \eqref{smallf} as
\begin{align}
|f_{\chi,\text{ tree}}|\sim 3\times 10^3 \times \frac{\rho_2^2}{\dot{\sigma}_0},
\end{align}
where we have denoted the quadratic mixing as $\rho_2 = \frac{2\dot{\sigma}_0v}{\Lambda_\sigma^2}$. Since $\dot{\sigma}_0\sim H^2$, compared to the result in the inflationary paradigm given in eq. \eqref{treehiggsng}, we can have orders of magnitude larger NG in the curvaton scenario, and importantly, \textit{without} any classical tuning. 

\begin{figure}[h]
	\centering
	\includegraphics[width=0.6\linewidth]{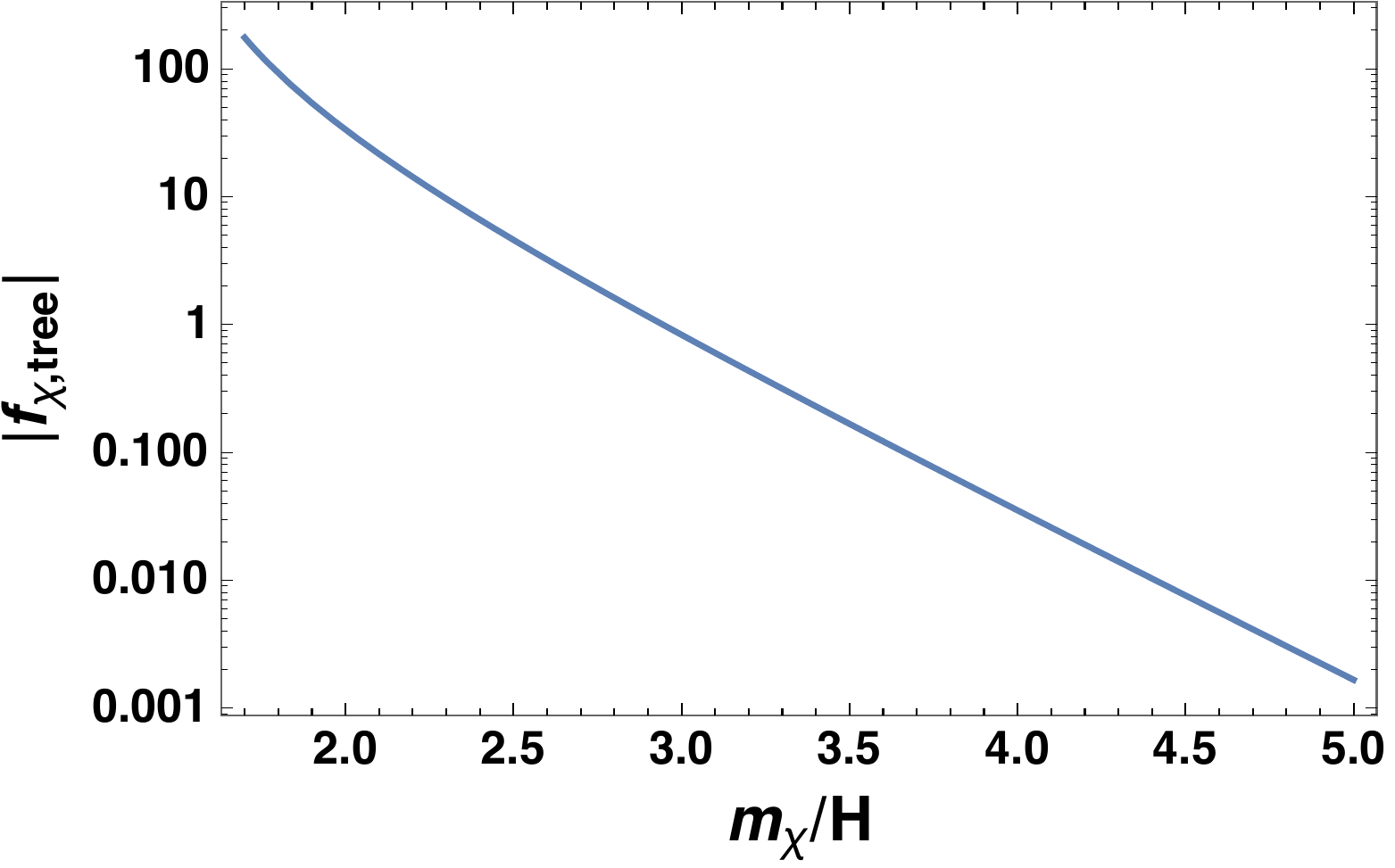}
	\caption{The strength of NG for tree level Higgs exchange as a function of Higgs mass $m_\chi$ for $\rho_2=0.3H$ and $\dot{\sigma}_0=-H^2$. The function $|f_{\chi,\text{tree}}(\mu)|$ is defined in eq. \eqref{fscalartree}.}
	\label{fig:tree-higgs}
\end{figure}

While the above is a rough estimate, the necessary ingredients for a precise calculation of the single exchange diagram in fig. \ref{fig:diagrams}, can be found in \cite{Arkani-Hamed:2015bza} using which we get,
\begin{align}\label{fscalartree}
&F_{\chi,\text{tree}}\left(m_\chi,\frac{k_3}{k_1}\right)=\frac{\langle \zeta(\vec{k}_1) \zeta(\vec{k}_2)\zeta(\vec{k}_3)\rangle^\prime}{\langle \zeta(\vec{k}_1) \zeta(-\vec{k}_1)\rangle^\prime \langle \zeta(\vec{k}_3) \zeta(-\vec{k}_3)\rangle^\prime}\nonumber\\
&=-\frac{3\sigma_0 \rho_2^2}{8\dot{\sigma}_0 H}
\left(\Gamma(\frac{1}{2} + i \mu)^2 \Gamma(-2 i \mu) (\frac{3}{2} + i \mu) (\frac{5}{2} + 
i \mu) (1 + i \sinh(\pi \mu)) \left(\frac{k_3}{k_1}\right)^{\frac{3}{2} + i \mu} + \mu\rightarrow -\mu\right)\nonumber \\
&\equiv |f_{\chi,\text{tree}}(\mu)| \left(e^{i\delta_1(\mu)}\left(\frac{k_3}{k_1}\right)^{\frac{3}{2} + i \mu}+\mu\rightarrow-\mu\right),
\end{align}
where $\mu=\sqrt{m_\chi^2/H^2-9/4}$. In fig. \ref{fig:tree-higgs} we plot the function $|f_{\chi,\text{tree}}|$ which gives the strength of NG as a function of the scalar mass, $m_\chi$.

\subsection{Charged scalar exchange in the symmetric phase}
The leading coupling is again given by a similar dimension-6 operator,
\begin{equation}\label{curv-higgs}
    \mathcal{L}\supset\frac{1}{\Lambda_\sigma^2}(\partial\sigma)^2 \chi^\dagger\chi.
  \end{equation}
In the absence of symmetry breaking, the relevant curvaton-$\chi$ interaction terms are given by,
\begin{equation}\label{curv-Higgs-coupling}
    \left( - \frac{2\dot{\sigma}_0}{\Lambda_\sigma^2} \dot{\delta\sigma}+\frac{1}{\Lambda_\sigma^2}(\partial(\delta\sigma))^2\right)\chi^\dagger\chi. 
\end{equation}
Using eq. \eqref{curv-Higgs-coupling} we can estimate the strengths of NG. The relevant diagrams are still given by fig. \ref{fig:scalar-loop} except the external legs now represent curvaton fluctuations instead of inflaton fluctuations. The leading NG is given by the double exchange diagram in fig. \ref{fig:scalar-loop} whose 
parametric strength, in terms of $f$ defined eq. \eqref{smallf}, is given by,
\begin{equation}
|f_{\chi,\text{loop}}|\sim \frac{1}{16\pi^2}\frac{\dot{\sigma}_0 H^2}{\Lambda_\sigma^4}\times\frac{\sigma_0}{H}.
\end{equation}  
Since $\Lambda_\sigma \ll \sqrt{\dot{\phi}_0}$, compared to the result in the inflationary paradigm given in eq. \eqref{brokenscalrinf}, we can have orders of magnitude bigger NG in the curvaton scenario, and again, \textit{without} any classical tuning. 

\begin{figure}[h]
	\centering
	\includegraphics[width=0.6\linewidth]{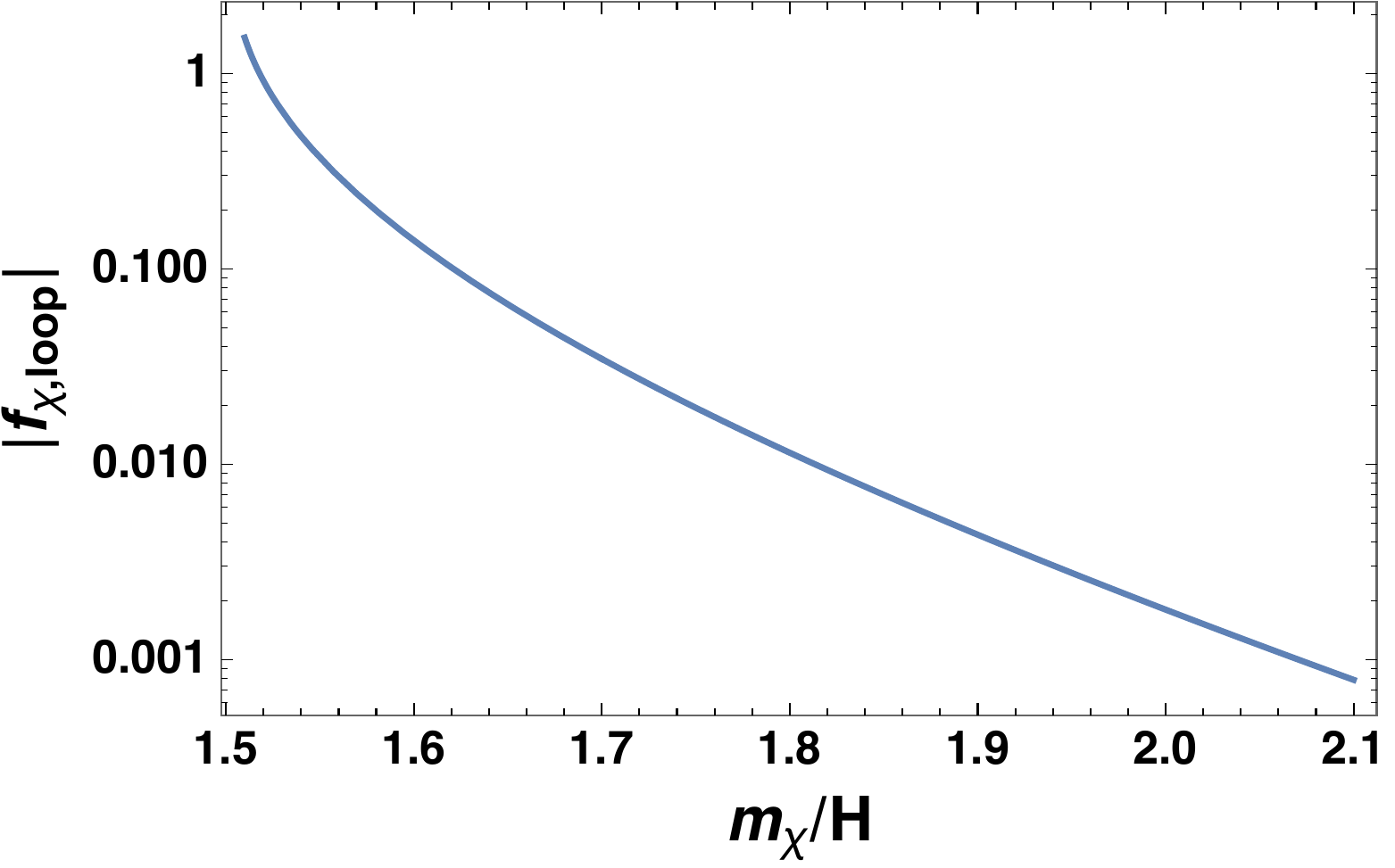}
	\caption{The strength of NG for loop level scalar exchange as a function of scalar mass $m_\chi$ for $\Lambda_{\sigma}=4H$ and $\dot{\sigma}_0=-H^2$. The function  $|f_{\chi,\text{loop}}(\mu)|$ is defined in eq.  \eqref{fscalarloop}.}
	\label{fig:loop-higgs}
\end{figure}

The dimensionless three point function, defined in eq. \eqref{F}, due to the double exchange diagram in fig. \ref{fig:scalar-loop} will be calculated in appendix \ref{scalarloop}. The result is given by eq. \eqref{scalarloopF},
\begin{align}\label{fscalarloop}
F_{\chi,\text{loop}}\left(m_\chi,\frac{k_3}{k_1}\right)&=\frac{\langle \zeta(\vec{k}_1) \zeta(\vec{k}_2)\zeta(\vec{k}_3)\rangle^\prime}{\langle \zeta(\vec{k}_1) \zeta(-\vec{k}_1)\rangle^\prime \langle \zeta(\vec{k}_3) \zeta(-\vec{k}_3)\rangle^\prime}\nonumber\\
&=-\frac{3\sigma_0}{2}\frac{2\dot{\sigma}_0 H}{\Lambda_\sigma^4} \left( \frac{\frac{1}{16\pi^5}\Gamma(-i\mu)^2\Gamma(3/2+i\mu)^2}{\frac{1}{4\pi^{5/2}}\Gamma(-3/2-2i\mu)\Gamma(3+2i\mu)}\mathcal{F}\left(3+2i\mu,\frac{k_3}{k_1}\right)+\mu\rightarrow -\mu\right)\nonumber \\
&\equiv |f_{\chi,\text{loop}}(\mu)| \left(e^{i\delta_2(\mu)}\left(\frac{k_3}{k_1}\right)^{3+2i\mu}+\mu\rightarrow-\mu\right),
\end{align}
where $\mathcal{F}$ is defined in eq. \eqref{curlyF} and $\mu=\sqrt{m_\chi^2/H^2-9/4}$.
In fig. \ref{fig:loop-higgs} we plot the function $|f_{\chi,\text{loop}}(\mu)|$ as a function of the scalar mass, $m_\chi$. 

\subsection{Charged Dirac fermion}
We again consider a charged Dirac fermion, and its coupling to the curvaton is given by a similar dimension-7 operator,
\begin{equation}
    \frac{1}{\Lambda_\sigma^3}(\partial\sigma)^2\bar{\Psi}\Psi.
\end{equation}
As before, there can be a dimension-5 axial coupling $\frac{\partial_\mu\sigma\bar{\Psi}\gamma^\mu\gamma^5\Psi}{\Lambda_\sigma}$ between $\sigma$ and $\Psi$ if it is not forbidden by parity. However, such a coupling does not give rise to a significant chemical potential as in the inflationary scenario, and correspondingly there is no substantial enhancement of NG signals. We leave a detailed study of this coupling for future work, focusing only on the dimension-7 operator in the present work.

Through a reasoning identical to the case of scalars, one can check that the mass correction to the fermion due to the curvaton coupling is negligible with similar choice of parameters as above. The relevant curvaton-fermion interaction terms are given by,
\begin{equation}\label{curv-fermion-coupling}
    \left( - \frac{2\dot{\sigma}_0}{\Lambda_\sigma^3} \dot{\delta\sigma}+\frac{1}{\Lambda_\sigma^3}(\partial(\delta\sigma))^2\right)\bar{\Psi}\Psi.
\end{equation}
Using eq. \eqref{curv-fermion-coupling} we can estimate the strengths of NG for which the relevant diagrams are still given by fig. \ref{fig:scalar-loop}, except the external legs now represent curvaton fluctuations instead of inflaton fluctuations. The leading NG is given by the double exchange diagram in fig. \ref{fig:scalar-loop} whose 
parametric strength, in terms of $f$ defined eq. \eqref{smallf}, is given by,
\begin{equation}
    f_{\Psi,\text{loop}}\sim \frac{1}{16\pi^2}\frac{\dot{\sigma}_0 H^4}{\Lambda_\sigma^6}\times \frac{\sigma_0}{H}\vert_*.
\end{equation}  
Since $\Lambda_\sigma \ll \sqrt{\dot{\phi}_0}$, compared to the result in the inflationary paradigm given in eq. \eqref{loopfermion}, once again we can have orders of magnitude bigger NG in the curvaton scenario \textit{without} any classical tuning. 

\begin{figure}[h]
	\centering
	\includegraphics[width=0.6\linewidth]{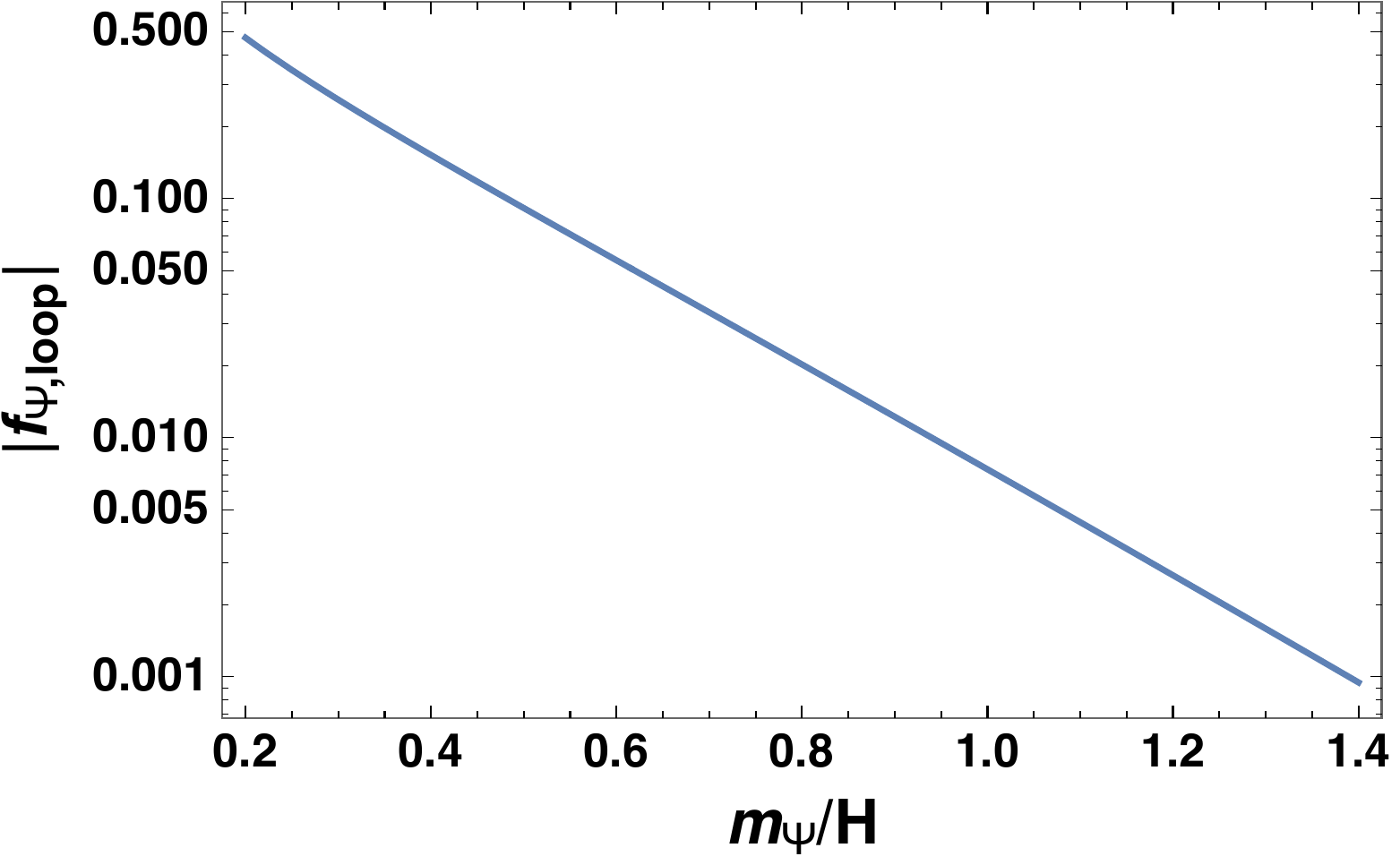}
	\caption{The strength of NG for loop level charged fermion exchange as a function of fermion mass $m_\Psi$ for $\Lambda_{\sigma}=4H$ and $\dot{\sigma}_0=-H^2$. The function  $|f_{\Psi,\text{loop}}(\tilde{\mu})|$ is defined in eq.  \eqref{ffermionloop}.}
	\label{fig:loop-fermion}
\end{figure}

The dimensionless three point function, defined in eq. \eqref{F}, due to the double exchange diagram in fig. \ref{fig:scalar-loop} will be calculated in appendix \ref{fermionloop}. The result is given by eq. \eqref{fermionloopF},
\begin{align}\label{ffermionloop}
F_{\Psi,\text{loop}}\left(m_\Psi,\frac{k_3}{k_1}\right)&=\frac{\langle \zeta(\vec{k}_1) \zeta(\vec{k}_2)\zeta(\vec{k}_3)\rangle^\prime}{\langle \zeta(\vec{k}_1) \zeta(-\vec{k}_1)\rangle^\prime \langle \zeta(\vec{k}_3) \zeta(-\vec{k}_3)\rangle^\prime}\nonumber\\
&=-\frac{3\sigma_0}{2}\frac{2\dot{\sigma}_0H^3}{\Lambda_\sigma^6} \left( \frac{-\frac{3}{\pi^5}\frac{\Gamma(1/2-i\tilde{\mu})^2\Gamma(2+i\tilde{\mu})^2}{(1+2i\tilde{\mu})}}{\frac{1}{4\pi^{5/2}}\Gamma(-5/2-2i\tilde{\mu})\Gamma(4+2i\tilde{\mu})}\mathcal{F}\left(4+2i\tilde{\mu},\frac{k_3}{k_1}\right)+\tilde{\mu}\rightarrow -\tilde{\mu}\right)\nonumber\\
&\equiv |f_{\Psi,\text{loop}}(\tilde{\mu})| \left(e^{i\delta_3(\tilde{\mu})}\left(\frac{k_3}{k_1}\right)^{4+2i\tilde{\mu}}+\tilde{\mu}\rightarrow-\tilde{\mu}\right),
\end{align}
where $\mathcal{F}$ is defined in eq. \eqref{curlyF} and $\tilde{\mu} = m_\Psi/H$.
In fig. \ref{fig:loop-fermion} we plot the function $|f_{\Psi,\text{loop}}|$ as a function of fermion mass, $m_\Psi$.

\section{Conclusions and future directions}\label{conclude}
In the standard inflationary scenario, the inflaton-heavy particle couplings are suppressed by (at least) inverse powers of an EFT cutoff scale $\Lambda_{\phi}$, which has to satisfy $\Lambda_\phi\gtrsim V_{\text{inf}}^{\frac{1}{4}}>250 H$ if we demand that the EFT explicitly describes the scalar field source of inflation $V_{\text{inf}}$. Such a high value of $\Lambda_\phi$ leads to small or even unobservable strengths of NG in various otherwise well-motivated scenarios of cosmological collider physics.

We have studied a non-standard inflationary paradigm in which the space-time expansion and the predominant production of primordial fluctuations are due to two \textit{different} fields. In particular, we have focused on the curvaton scenario, where the above two roles are played by the inflaton and the curvaton respectively. Heavy particles can then couple to the curvaton suppressed only by inverse powers of $\Lambda_\sigma\gtrsim V_{\sigma}^{\frac{1}{4}}$, where $V_\sigma$ is the energy density contained in the curvaton field which can be as low as $\sim (10 H)^4$ for the benchmark parameter choice considered in this work. With the choice $V_{\sigma}^{\frac{1}{4}}\lesssim\Lambda_\sigma \ll \Lambda_\phi$, the heavy particles can couple to the primordial fluctuations much more strongly and lead to orders of magnitude larger NG while still ensuring controlled EFT description of the inflationary and reheating dynamics. To illustrate this fact, we have considered NG mediated by charged scalars, both in the Higgs phase and in the unbroken phase, as well as charged Dirac fermions. In particular, we have shown that even loop-level NG effects are observable!

Several future directions remain open. We have considered a scenario in which the curvaton and the inflaton belong to two decoupled sectors (up to gravitational effects), which can naturally be achieved, for example, by having an extra dimension in which the two fields are localized on two different branes. Since the presence of Kaluza-Klein (KK) gravitons is a robust feature of such an extra-dimensional set-up, it would be interesting to see whether appreciable KK-graviton mediated NG can be generated using our scenario. In particular, this may be applicable to the strongly motivated case of orbifold unification \cite{Kawamura:1999nj,Kawamura:2000ev,Hall:2002ea} along the lines of \cite{Kumar:2018jxz}.

We have restricted our attention to bispectra of primordial scalar fluctuations which required us to use a non-zero value of $\dot{\sigma}_0$ in the ``in-in'' diagrams. Demanding $\dot{\sigma}_0\sim H^2$ then implied $V_\sigma\sim (10H)^4$ via eqs. \eqref{Vsigma}, \eqref{sigmadot} and \eqref{sigma0}. Considering trispectra (four-point correlation functions) on the other hand, such insertions of $\dot{\sigma}_0$ are not essential, and consequently one can have even smaller values of $V_\sigma$. This will then allow smaller $\Lambda_\sigma$ and larger NG. It would also be very interesting to analyze all the ``heavy-lifted'' SM signals \cite{Chen:2016hrz,Chen:2016nrs,Chen:2016uwp,Kumar:2017ecc}, especially the signal due to the loops of massive $W$ bosons, in the curvaton scenario that was presented in this work. 

\section*{Acknowledgements}
The authors would like to thank Anson Hook for helpful discussions. This research was supported in part by the NSF grants PHY-1620074 and PHY-1914731, and by the Maryland Center for Fundamental Physics (MCFP). RS acknowledges the hospitality of the Kavli Institute for Theoretical Physics, UC Santa Barbara, during the ``Origin of the Vacuum Energy and Electroweak Scales'' workshop, and the support by the NSF grant PHY-174958.

\appendix
\section{Charged scalar loop}\label{scalarloop}
The three point function induced by an interaction of the type $g (\partial\phi)^2\mathcal{O}$,  where $\mathcal{O}$ could be either an elementary or a composite operator, was calculated in \cite{Arkani-Hamed:2015bza}. This was done by evaluating the coefficients $c_\Delta$ and the scaling dimensions $\Delta$ which appear in the late-time limit (i.e. $\eta,\eta^\prime\rightarrow 0$) of the position space two point correlation function:
\begin{equation}\label{generaltwopoint}
\langle \mathcal{O}(\eta,\vec{x})\mathcal{O}(\eta^\prime,\vec{x}^\prime)\rangle = \sum_\Delta c_\Delta \left(\frac{\eta\eta^\prime}{|\vec{x}-\vec{x}^\prime|}\right)^\Delta.
\end{equation}
Given $c_\Delta$'s for the set of $\Delta$'s, the three point function can be written as \cite{Arkani-Hamed:2015bza},
\begin{equation}\label{inflatonthreept}
    \langle \delta\phi(\vec{k}_1) \delta\phi(\vec{k}_2) \delta\phi(\vec{k}_3)\rangle^\prime = -\frac{g^2\dot{\phi}_0}{2}\frac{1}{k_1^3k_3^3}\sum_\Delta \frac{c_\Delta}{c_\text{free}(\Delta)}\mathcal{F}\left(\Delta,\frac{k_3}{k_1}\right),
\end{equation}
where
\begin{equation}\label{cfree}
    c_\text{free}(\Delta) = \frac{1}{4\pi^{5/2}}\Gamma(3/2-\Delta)\Gamma(\Delta),
\end{equation}
and
\begin{equation}\label{curlyF}
\mathcal{F}\left(\Delta,\frac{k_3}{k_1}\right) = \frac{4^{-\Delta+3/2}\pi^{3/2}}{4\cos(\pi(\Delta-3/2))^2}\frac{(1+\sin(\pi(\Delta-3/2)))\Delta (\Delta+1)\Gamma(3/2-\Delta)}{\Gamma(2-\Delta)}\left(\frac{k_3}{k_1}\right)^\Delta.
\end{equation}
In this and the following appendix, we will work in the units where $H=1$.

\begin{figure}[h]
	\centering
	\includegraphics[width=0.7\linewidth]{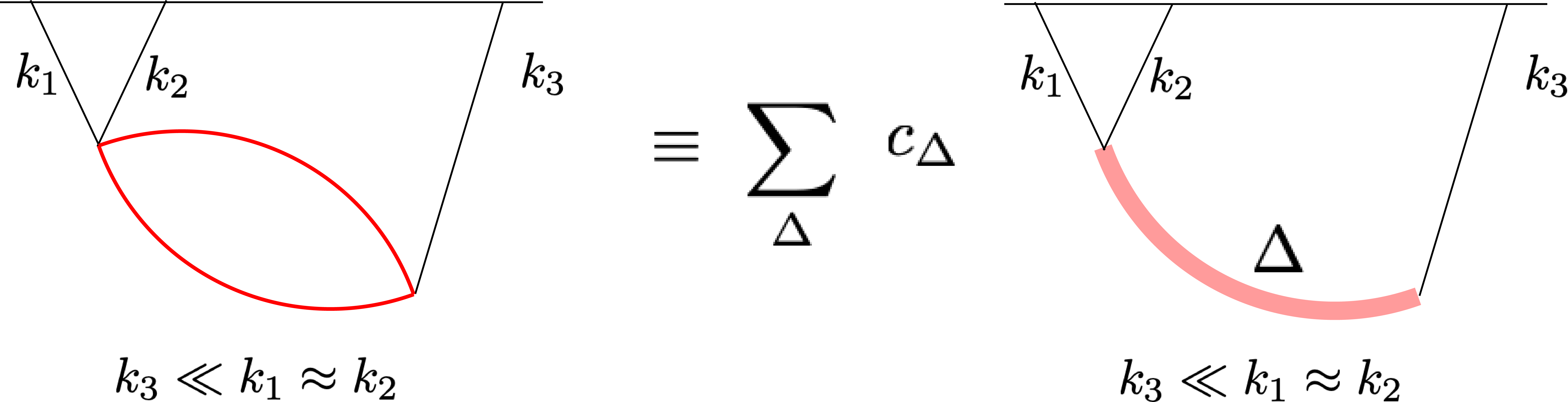}
	\caption{Reduction of a loop diagram into a linear combination of tree diagrams in the squeezed limit.}
	\label{fig:loop-ope}
\end{figure}

Now we are ready to give the three point function induced by a Higgs loop. To get the associated $c_\Delta$'s we first write,
\begin{equation}\label{wick}
\langle \chi^\dagger \chi(\eta,\vec{x})\cdot \chi^\dagger \chi(\eta^\prime,\vec{x}^\prime)\rangle  = \langle \chi_1(\eta,\vec{x})\chi_1(\eta^\prime,\vec{x}^\prime)\rangle^2,  
\end{equation}
where we have written $\chi = \frac{1}{\sqrt{2}}(\chi_1+i\chi_2)$ and used the fact that $\langle \chi_1(\eta,\vec{x}) \chi_1(\eta^\prime,\vec{x}^\prime)\rangle  = \langle \chi_2(\eta,\vec{x}) \chi_2(\eta^\prime,\vec{x}^\prime)\rangle$. Using the fact that the non-analytic pieces of the two point function are given by,
\begin{equation}\label{scalarlatetime}
\langle \chi_1(\eta,\vec{x})\chi_1(\eta^\prime,\vec{x}^\prime)\rangle^2\vert_{\eta,\eta^\prime\rightarrow 0} = \frac{1}{16\pi^5}\left(\left(\frac{\eta\eta^\prime}{|\vec{x}-\vec{x}^\prime|^2}\right)^{3+2i\mu}\Gamma(-i\mu)^2\Gamma(3/2+i\mu)^2 + \mu\rightarrow -\mu \right),
\end{equation}
where $\mu=\sqrt{m_\chi^2-9/4}$, we see that the composite operator $\chi^\dagger \chi$ gives rise to scaling dimensions $\Delta=3\pm 2i\mu,3$. 
More generally, in the squeezed limit, a loop diagram can be decomposed into a set of effective ``tree'' diagrams, each involving a dS mass eigenstate corresponding to such a scaling dimension $\Delta$ \cite{Arkani-Hamed:2015bza}. This is schematically shown in fig. \ref{fig:loop-ope}.
Thus from eqs. \eqref{generaltwopoint}, \eqref{cfree} and \eqref{scalarlatetime}, we will have,
\begin{align}
c_{3+2i\mu} &= \frac{1}{16\pi^5}\Gamma(-i\mu)^2\Gamma(3/2+i\mu)^2,\\
c_\text{free}(3+2i\mu) &= \frac{1}{4\pi^{5/2}}\Gamma(-3/2-2i\mu)\Gamma(3+2i\mu),
\end{align}
which can be used to get the inflaton three point function in eq. \eqref{inflatonthreept}. 
We can quickly generalize this to the case of curvaton fluctuations by using eq. \eqref{curv-Higgs-coupling} to get,
\begin{align}
\langle \delta\sigma(\vec{k}_1) \delta\sigma(\vec{k}_2)& \delta\sigma(\vec{k}_3)\rangle^\prime \nonumber\\&= -\frac{\dot{\sigma}_0}{2\Lambda_\sigma^4}\frac{1}{k_1^3k_3^3}\left( \frac{\frac{1}{16\pi^5}\Gamma(-i\mu)^2\Gamma(3/2+i\mu)^2}{\frac{1}{4\pi^{5/2}}\Gamma(-3/2-2i\mu)\Gamma(3+2i\mu)}\mathcal{F}\left(3+2i\mu,\frac{k_3}{k_1}\right)+\mu\rightarrow -\mu\right).
\end{align}
Using the above and eq. \eqref{zetafin}, we derive the dimensionless three point function of the curvature perturbation,
\begin{align}\label{scalarloopF}
F_{\chi,\text{loop}}\left(m_\chi,\frac{k_3}{k_1}\right)&=\frac{\langle \zeta(\vec{k}_1) \zeta(\vec{k}_2)\zeta(\vec{k}_3)\rangle^\prime}{\langle \zeta(\vec{k}_1) \zeta(-\vec{k}_1)\rangle^\prime \langle \zeta(\vec{k}_3) \zeta(-\vec{k}_3)\rangle^\prime}\nonumber\\
&=-\frac{3\sigma_0}{2}\frac{2\dot{\sigma}_0}{\Lambda_\sigma^4} \left( \frac{\frac{1}{16\pi^5}\Gamma(-i\mu)^2\Gamma(3/2+i\mu)^2}{\frac{1}{4\pi^{5/2}}\Gamma(-3/2-2i\mu)\Gamma(3+2i\mu)}\mathcal{F}\left(3+2i\mu,\frac{k_3}{k_1}\right)+\mu\rightarrow -\mu\right)\nonumber \\
&\equiv |f_{\chi,\text{loop}}(\mu)| \left(e^{i\delta_2(\mu)}\left(\frac{k_3}{k_1}\right)^{3+2i\mu}+\mu\rightarrow-\mu\right).
\end{align}
\section{Fermion loop}\label{fermionloop}
To calculate the NG induced by the coupling given in eq. \eqref{curv-fermion-coupling}, we need the late time two point function of the type $\langle \bar{\Psi} \Psi(\eta,\vec{x})\cdot \bar{\Psi} \Psi(\eta^\prime,\vec{x}^\prime)\rangle$. This can be calculated by squaring and taking a trace of the spinor two point function $\langle  \Psi(\eta,\vec{x}) \bar{\Psi}(\eta^\prime,\vec{x}^\prime)\rangle$ derived in \cite{allen1986}. The result is,
\begin{align}
\langle \bar{\Psi} \Psi(\eta,\vec{x}) \bar{\Psi} \Psi(\eta^\prime,\vec{x}^\prime)\rangle \vert_{\eta,\eta^\prime\rightarrow 0} \nonumber &\\=
& -\frac{3}{\pi^5}\left(\frac{\eta\eta^\prime}{|\vec{x}-\vec{x}^\prime|^2}\right)^{4+2i\mu} \frac{\Gamma(1/2-i\tilde{\mu})^2\Gamma(2+i\tilde{\mu})^2}{(1+2i\tilde{\mu})} + \tilde{\mu}\rightarrow-\tilde{\mu},
\end{align}
where $\tilde{\mu}=m_\Psi$. This matches with the answer obtained in \cite{Lu:2019tjj}. We thus get using eqs. \eqref{generaltwopoint} and \eqref{cfree},
\begin{align}
c_{4+2i\tilde{\mu}} = -\frac{3}{\pi^5}\frac{\Gamma(1/2-i\tilde{\mu})^2\Gamma(2+i\tilde{\mu})^2}{(1+2i\tilde{\mu})},\\
c_{\text{free}}(4+2i\tilde{\mu}) = \frac{1}{4\pi^{5/2}}\Gamma(-5/2-2i\tilde{\mu})\Gamma(4+2i\tilde{\mu}). 
\end{align}
Using eq. \eqref{curv-fermion-coupling} we can get the three point function of the curvaton fluctuations,
\begin{align}
\langle \delta\sigma(\vec{k}_1) \delta\sigma(\vec{k}_2)& \delta\sigma(\vec{k}_3)\rangle^\prime \nonumber\\&= -\frac{\dot{\sigma}_0}{2\Lambda_\sigma^6}\frac{1}{k_1^3k_3^3}\left( \frac{-\frac{3}{\pi^5}\frac{\Gamma(1/2-i\tilde{\mu})^2\Gamma(2+i\tilde{\mu})^2}{(1+2i\tilde{\mu})}}{\frac{1}{4\pi^{5/2}}\Gamma(-5/2-2i\tilde{\mu})\Gamma(4+2i\tilde{\mu})}\mathcal{F}\left(4+2i\tilde{\mu},\frac{k_3}{k_1}\right)+\tilde{\mu}\rightarrow -\tilde{\mu}\right).
\end{align}
Using the above and eq. \eqref{zetafin}, the dimensionless bispectrum is given by,
\begin{align}\label{fermionloopF}
F_{\Psi,\text{loop}}\left(m_\Psi,\frac{k_3}{k_1}\right)&=\frac{\langle \zeta(\vec{k}_1) \zeta(\vec{k}_2)\zeta(\vec{k}_3)\rangle^\prime}{\langle \zeta(\vec{k}_1) \zeta(-\vec{k}_1)\rangle^\prime \langle \zeta(\vec{k}_3) \zeta(-\vec{k}_3)\rangle^\prime}\nonumber\\
&=-\frac{3\sigma_0}{2}\frac{2\dot{\sigma}_0}{\Lambda_\sigma^6} \left( \frac{-\frac{3}{\pi^5}\frac{\Gamma(1/2-i\tilde{\mu})^2\Gamma(2+i\tilde{\mu})^2}{(1+2i\tilde{\mu})}}{\frac{1}{4\pi^{5/2}}\Gamma(-5/2-2i\tilde{\mu})\Gamma(4+2i\tilde{\mu})}\mathcal{F}\left(4+2i\tilde{\mu},\frac{k_3}{k_1}\right)+\tilde{\mu}\rightarrow -\tilde{\mu}\right)\nonumber\\
&\equiv |f_{\Psi,\text{loop}}(\tilde{\mu})| \left(e^{i\delta_3(\tilde{\mu})}\left(\frac{k_3}{k_1}\right)^{4+2i\tilde{\mu}}+\tilde{\mu}\rightarrow-\tilde{\mu}\right).
\end{align}

\bibliographystyle{JHEP}

\bibliography{refs}

 \end{document}